\documentclass[showpacs,preprintnumbers,superscriptaddress,prl,twocolumn]{revtex4-1}

\usepackage{graphicx}
\usepackage{amsmath}
\usepackage{times}

\newcommand{\im}{i}
\newcommand{\ex}{e}

\newcommand{\therm}{\mathrm{th}}

\begin{document}

\title[Scheme for steady-state preparation of a harmonic oscillator in the first excited state]{Scheme for steady-state preparation of a harmonic oscillator in the first excited state}
\author{Kjetil B{\o}rkje}
\affiliation{Niels Bohr Institute, University of Copenhagen, Blegdamsvej 17, DK-2100 Copenhagen, Denmark}

\date{\today}

\begin{abstract}
We present a generic quantum master equation whose dissipative dynamics autonomously stabilizes a harmonic oscillator in the $n=1$ Fock state. A multi-mode optomechanical system is analyzed and shown to be an example of a physical system obeying this model. We show that the optomechanical setup enables preparation of a mechanical oscillator in a nonclassical steady state, and that this state indeed approaches a single phonon Fock state in the ideal parameter regime. The generic model may be useful in other settings, such as cavity or circuit quantum electrodynamics or trapped ion physics.
\end{abstract}


\maketitle

\emph{Introduction.}
Dissipation and decoherence from unwanted interactions with the environment are universal obstacles when trying to manipulate and control quantum systems. In some cases, a quantum state can be sustained beyond the timescale set by coupling to the environment by utilizing measurement-based feedback schemes \cite{Sayrin2011Nature,Vijay2012Nature,Riste2012PRL,Campagne-Ibarcq2013PRX}. 
The interaction with the environment can occasionally even be used to ones advantage in order to stabilize a desired quantum state - a concept known as {\it quantum reservoir engineering} \cite{Poyatos1996PRL}. This involves designing the experiment such that the steady state of the system in presence of dissipation and decoherence equals the desired state, which eliminates the need for an active feedback scheme. This concept has been used in experiments both to stabilize the state of a single qubit \cite{Geerlings2013PRL,Murch2012PRL} and to prepare two qubits in an entangled steady state \cite{Lin2013Nature,Shankar2013Nature}.

%

In this article, we present a new quantum reservoir engineering scheme. We study a generic model whose dissipative dynamics in a certain parameter regime prepares a harmonic oscillator in the $n=1$ number state. As an example of a physical system described by this model, we analyze an optomechanical system \cite{Aspelmeyer2013} where a mechanical oscillator is parametrically coupled to several optical cavity modes. We show that the quantum master equation describing this system can be mapped onto the generic model. By solving the full quantum master equation for the optomechanical system numerically, we demonstrate that the mechanical oscillator relaxes into a nonclassical state, characterized by negativity of the Wigner quasi-probability distribution. This occurs despite the fact that the mechanical oscillator is in contact with a thermal reservoir. Furthermore, we show that for ideal parameters, this nonclassical state approaches the $n=1$ Fock state. We note that another optomechanical scheme for single phonon Fock state preparation has previously been proposed \cite{Rips2012NJP}, but that required an intrinsic mechanical nonlinearity which is not necessary in our scheme. 

The optomechanical example we study requires that the single-photon optomechanical coupling rate \cite{Aspelmeyer2013} exceeds the intrinsic decay rates of the optical cavity modes and the mechanical oscillator. This has been realized in experiments where the mechanical element is a cloud of cold atoms \cite{Murch2008NatPhys}, but these experiments suffer from a very small mechanical frequency. The desired regime may however be within reach for other optomechanical realizations, e.g.~with optomechanical crystals \cite{Safavi-Naeini2011NJP,Davanco2012OptExp}, or superconducting circuits \cite{Teufel2011Nature_2}. 
Theoretical studies of this regime have investigated the optical cavity response \cite{Borkje2013PRL,Lemonde2013PRL,Kronwald2013PRL,Liu2013PRL}, prospects of quantum nondemolition measurements of phonon and photon numbers \cite{Ludwig2012PRL}, and the possibility of unitary quantum gate operations \cite{Stannigel2012PRL}. It has been reported that mechanical steady states with negative Wigner distributions can appear in the regime where self-sustained mechanical oscillations take place \cite{Qian2012PRL,Lorch2014PRX,Rodrigues2010PRL,Nation2013PRA}. We show that this is also possible when the oscillator is not undergoing coherent oscillations. Moreover, this article is to our knowledge the first to propose how a specific mechanical steady state with a negative Wigner distribution can be engineered using only the nonlinearity of the three-wave mixing radiation pressure interaction and simple continuous optical driving.
%

We also speculate that our generic model can be useful for realizing Fock states in other physical systems, such as photon number states with cavity or circuit quantum electrodynamics (cQED) or phonon number states with trapped ions. We note that a different stabilization scheme for photon number states have been proposed in the context of cQED \cite{Sarlette2012}. 

\emph{Generic model}. We consider a quantum master equation $\dot{\hat{\rho}} = \left({\cal L}_c + {\cal L}_d\right) \hat{\rho}$, where $\hat{\rho}$ is the system density matrix. The coherent part of the Liouvillian is given by
\begin{eqnarray}
\label{eq:Lc}
{\cal L}_c\hat{\rho} & = & - i \tilde{g} \left[\hat{a}^\dagger \hat{c}^2 + \hat{c}^{\dagger \, 2} \hat{a} \, ,\hat{\rho}\right] ,
\end{eqnarray}
where $\hat{c}$ is the annihilation operator for a harmonic oscillator in a frame rotating at its resonance frequency. The operator $\hat{a}$ can for example be an annihilation operator for another harmonic oscillator, or a spin lowering operator for a two-level system ($\hat{a} \rightarrow \hat{\sigma}_-$). The dissipative part of the Liouvillian is
\begin{eqnarray}
\label{eq:Ld}
{\cal L}_d & = & \Gamma {\cal D}[\hat{c}^\dagger \hat{a}] + \kappa {\cal D}[\hat{a}] + \gamma_\downarrow {\cal D}[\hat{c}]  + \gamma_\uparrow {\cal D}[\hat{c}^\dagger]  ,
\end{eqnarray}
with ${\cal D}[\hat{o}]\hat{\rho} = \hat{o} \hat{\rho} \hat{o}^\dagger - \left(\hat{o}^\dagger \hat{o} \hat{\rho} + \hat{\rho} \hat{o}^\dagger \hat{o} \right)/2$ being the standard dissipator in Lindblad form. 

To analyze this model, let us first assume $\kappa = \gamma_\uparrow = \gamma_\downarrow = 0$ and that we start from a pure state $|0\rangle_{\hat{a}} \otimes |n\rangle_{\hat{c}}$, where $|0\rangle_{\hat{a}}$ is the ground state of system $\hat{a}$ and $|n\rangle_{\hat{c}}$ is a Fock state of oscillator $\hat{c}$ with $n > 1$. The interaction \eqref{eq:Lc} can then create an excitation in $\hat{a}$ by destroying two $\hat{c}$ particles, producing the state $|1\rangle_{\hat{a}} \otimes |n - 2\rangle_{\hat{c}}$. The excitation in $\hat{a}$ will be destroyed by the term proportional to $\Gamma$ in \eqref{eq:Ld}, which is accompanied by the creation of a $\hat{c}$ particle, giving the state $|0\rangle_{\hat{a}} \otimes |n - 1\rangle_{\hat{c}}$. We see that the excitation and subsequent deexcitation of $\hat{a}$ reduces the number of quanta in oscillator $\hat{c}$ by one, i.e.~it is a cooling process. However, this process only goes on until $n = 1$, in which case it stops. 

Including a nonzero $\kappa$ allows the excitation in $\hat{a}$ to be destroyed without the creation of a $\hat{c}$ particle. However, as long as $\kappa \ll \Gamma$, this process is suppressed. The term proportional to $\gamma_\downarrow$ ($\gamma_\uparrow$) describes processes where $\hat{c}$ particles are destroyed (created). If $\gamma_\uparrow \ll \mathrm{min}(4 \tilde{g}^2/\Gamma, \Gamma)$, the cooling process described above will nevertheless ensure that Fock states with $n > 1$ have negligible occupation. Additionally, if $\gamma_\downarrow \ll \gamma_\uparrow$, the occupation in state $|0\rangle_{\hat{c}}$ will be negligible compared to $|1\rangle_{\hat{c}}$. A more careful analysis \cite{SM} shows that in the right parameter regime, the steady state occupation probabilities $P_n$ of the harmonic oscillator Fock states $|n\rangle_{\hat{c}}$ obey $P_0/P_1 \approx \gamma_\downarrow/\gamma_\uparrow + 2\kappa/\Gamma$, $P_2/P_1 \approx \gamma_\uparrow \Gamma/(4 \tilde{g}^2) + 2 \gamma_\uparrow /\Gamma$, and $P_{n>2}/P_2 \ll 1$. This means that with the above assumptions, the oscillator $\hat{c}$ is approximately in the $n=1$ Fock state.

\begin{figure}[h]
 \centering
 \includegraphics[width=0.99\columnwidth,trim=0cm 0cm 0cm 0cm]{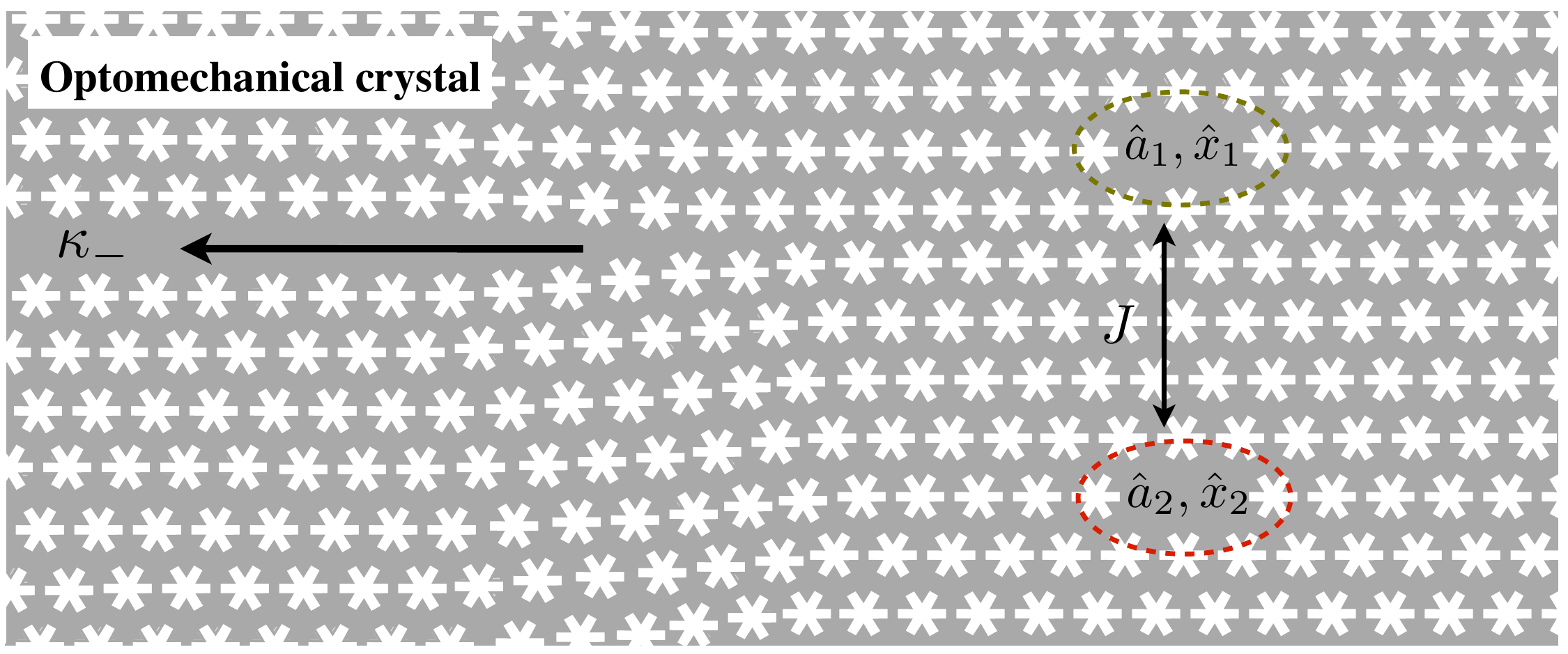}
\caption{(color online). Possible implementation. Two defects in a suspended two-dimensional photonic crystal give rise to co-localized optical and mechanical modes. An optical waveguide caused by a line defect provides external coupling to mode $\hat{a}_-$. The two mechanical modes $\hat{x}_1$ and $\hat{x}_2$ interact via phonon tunneling \cite{Safavi-Naeini2011NJP}, and $\hat{x}$ describes one of the two normal modes resulting from that interaction.}
\label{fig:Setup}
\end{figure}

\emph{The optomechanical system}.
We now move on to describe a physical system that can realize the generic model in Eqs.~\eqref{eq:Lc} and \eqref{eq:Ld}. We consider a system where two optical cavity modes are coupled to the same mechanical resonator. This could for example be realized in a two-dimensional optomechanical crystal \cite{Safavi-Naeini2011NJP} where co-localized optical and mechanical modes can be engineered, as depicted in Fig.~\ref{fig:Setup}. 
The system is described by the Hamiltonian
\begin{eqnarray}
\label{eq:H}
\hat{H} & = & \hbar \omega_1(\hat{x}) \hat{a}^\dagger_1 \hat{a}_1 + \hbar\omega_2(\hat{x}) \hat{a}^\dagger_2 \hat{a}_2 + \hbar \omega_m \hat{c}^\dagger \hat{c}  \\
& + & \hbar J \left( \hat{a}^\dagger_1 \hat{a}_2 + \hat{a}^\dagger_2 \hat{a}_1 \right) \ \notag
\end{eqnarray}
where for cavity $j = 1,2$, the resonance frequency is $\omega_j(\hat{x})$ and the photon annihilation operator is $\hat{a}_j$. The mechanical resonance frequency is $\omega_m$ and $\hat{c}$ is the annihilation operator for mechanical vibration quanta, i.e.~phonons. The mechanical displacement operator is $\hat{x} = x_0 + x_\mathrm{zpf} (\hat{c} + \hat{c}^\dagger)$, where $x_0 = \langle \hat{x} \rangle$ is the equilibrium position of the resonator and $x_\mathrm{zpf}$ the size of its zero point fluctuations. The second line in \eqref{eq:H} describes photon tunneling between the two optical modes, and we assume $|J| \ll \omega_1, \omega_2$. 

The interaction between the optical and mechanical degrees of freedom originates from the fact that the optical resonance frequencies depend parametrically on the position operator $\hat{x}$. To first order in $\hat{x} - x_0$, we have $\omega_j(\hat{x}) = \omega_j + (\partial \omega_j/\partial x)|_{x_0} x_\mathrm{zpf} (\hat{c} + \hat{c}^\dagger)$, where $\omega_j \equiv \omega_j(x_0)$. The Hamiltonian becomes $\hat{H} = \hat{H}_{\mathrm{free}} + \hat{H}_{\mathrm{int}}$, where the interaction Hamiltonian is $\hat{H}_{\mathrm{int}} = \hbar \left(\hat{c} + \hat{c}^\dagger \right) \left(g_1 \hat{a}^\dagger_1 \hat{a}_1 + g_2 \hat{a}^\dagger_2 \hat{a}_2 \right)$
and the absolute values of $g_j \equiv (\partial \omega_j/\partial x) |_{x_0} x_\mathrm{zpf}$ are the single-photon optomechanical coupling rates. 
  
\begin{figure}[h]
 \centering
 \includegraphics[width=0.99\columnwidth,trim=0cm 0cm 0cm 0cm]{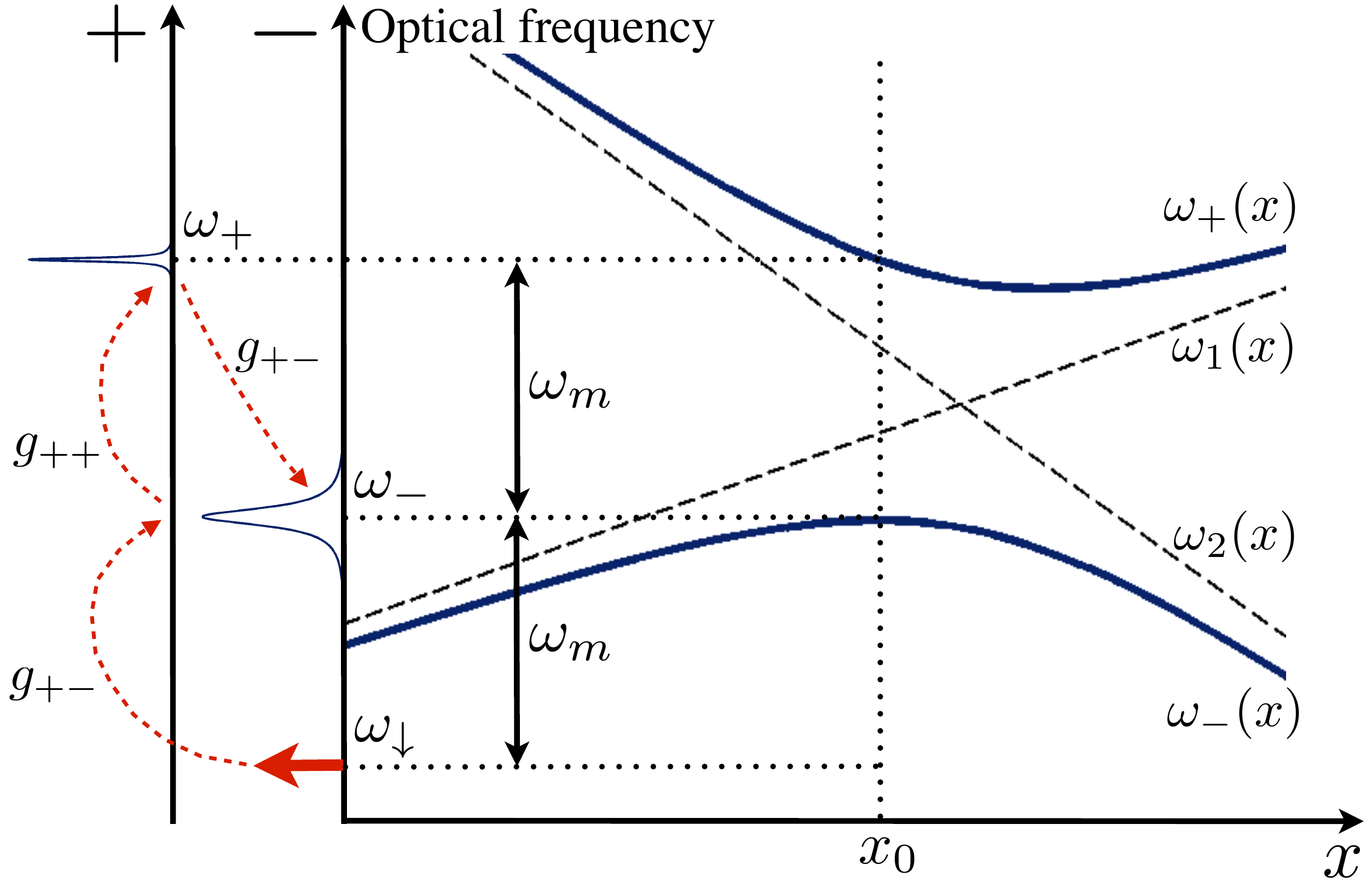}
\caption{(color online). Cavity resonance frequency as a function of position $x$.
Choosing the equilibrium position $x_0$ gives $g_{--} = 0$. On the left hand side, a three-step photon scattering process from the drive frequency $\omega_\downarrow$ to $\omega_-$ is depicted. The entire process destroys one phonon, but only occurs if the initial phonon number exceeds 1.}
\label{fig:Modes}
\end{figure}

\emph{Effective cavity modes}.  
Diagonalizing the free part of the Hamiltonian with a nonzero tunneling $J$ gives $\hat{H}_{\mathrm{free}} = \hbar \sum_{\mu = \pm} \omega_\mu \hat{a}^\dagger_\mu \hat{a}_\mu + \hbar \omega_m \hat{c}^\dagger \hat{c}$ where $\hat{a}_\pm$ are linear combinations of the original modes $\hat{a}_1$ and $\hat{a}_2$ \cite{SM} and the frequencies are $\omega_\pm = (\omega_1 + \omega_2)/2 \pm \sqrt{(\omega_2 - \omega_1)^2 + 4J^2}/2$. This gives rise to the anticrossing shown in Fig.~\ref{fig:Modes}. We will assume that $\omega_2 - \omega_1$ and $J$ are engineered in such a way that the mode splitting $\omega_+ - \omega_- \approx \omega_m$.

In terms of the new modes $\hat{a}_\pm$, the optomechanical interaction becomes $\hat{H}_{\mathrm{int}} = \hbar \left(\hat{c} + \hat{c}^\dagger \right) \sum_{\mu,\nu = \pm} g_{\mu \nu} \hat{a}^\dagger_\mu \hat{a}_\nu$. In general, we get both intramode ($g_{++}, g_{--}$) and intermode ($g_{+-} = g_{-+}$) optomechanical coupling in the new basis. The special case of $\omega_1 = \omega_2$ and $g_2 = -g_1$ gives only intermode coupling, and this case was studied in Refs.~\cite{Ludwig2012PRL,Stannigel2012PRL}. Here, we will choose the parameters such that $g_{--} = 0$ while $g_{++}$ and $g_{+-}$ are on the order of the original couplings $g_1$ and $g_2$. In the Supplementary Material \cite{SM}, we show that this is possible when assuming $\mathrm{sgn}(g_1 g_2) = -1$, $|g_1| \neq |g_2|$, and $\omega_1 \neq \omega_2$. In Fig.~\ref{fig:Modes}, choosing the equilibrium value $x_0$ indicated by the vertical dotted line leads to this situation, i.e. it gives $(\partial \omega_-/\partial x )|x_0 = 0$ and $(\partial \omega_+/\partial x )|x_0 \neq 0$. 

\emph{Dissipation and driving}. 
We let the energy decay rate of the effective cavity modes be $\kappa_+$ and $\kappa_-$, and make the natural assumption that $\hbar \omega_\pm \gg k_B T$ where $T$ is the temperature, such that the optical environment can be treated as zero temperature baths. Since the modes $\hat{a}_\pm$ both are combinations of $\hat{a}_1$ and $\hat{a}_2$ with comparable weights, one would expect $\kappa_+$ and $\kappa_-$ to be of the same order of magnitude \cite{SM}. However, we will assume that the two cavities are addressed via an optical waveguide, as illustrated in Fig.~\ref{fig:Setup}. If this waveguide is carefully positioned, one can selectively couple to only the mode $\hat{a}_-$ by exploiting destructive interference, and thereby increase the decay rate $\kappa_-$. We will also require that the optomechanical coupling strength exceeds the intrinsic dissipation rate, but not the one due to coupling to the waveguide, such that $\kappa_+ \ll |g_{++}|,|g_{+-}| \ll \kappa_-$. Finally, we assume that both the effective cavity modes satisfy the so-called resolved sideband condition $\omega_m \gg \kappa_\pm$.

The mode $\hat{a}_-$ can be coherently driven by utilizing the same waveguide as mentioned above. We let the drive frequency be $\omega_\downarrow = \omega_- - \omega_m$ (see Fig.~\ref{fig:Modes}). This adds a driving term $\hat{H}_\mathrm{drive} = \im \hbar (\Omega_\downarrow \ex^{-\im \omega_\downarrow t} \hat{a}_- - \mathrm{h.c.})$ to the Hamiltonian, where the rate $\Omega_\downarrow$ is proportional to the square root of the laser power.

The intrinsic mechanical energy decay rate will be denoted $\gamma$ and the temperature of the mechanical bath in units of quanta is $n_\therm = 1/(\ex^{\hbar \omega_m/k_B T} - 1)$. The physical temperature is of course always positive, which means $n_\therm > 0$. However, here we will create an apparent negative temperature bath by coupling the mechanical oscillator to a third cavity mode with annihilation operator $\hat{a}_3$. This is not such a demanding requirement, since optical cavities usually have many resonances, as do the devices depicted in Fig.~\ref{fig:Setup}. We require the linewidth $\kappa_3$ of this third optical mode to also be smaller than the mechanical frequency, but we can allow its optomechanical coupling $g_3$ to be weak, such that $\omega_m \gg \kappa_3 \gg |g_3|$. Note that it is not strictly necessary that $\kappa_3 \gg |g_3|$ for our scheme to work, but it is likely to be the case if the system has been engineered so as to maximize $|g_1|$ and $|g_2|$.  

If this third cavity mode is coherently driven at one mechanical frequency above the cavity resonance frequency $\omega_3$, i.e.~at $\omega_\uparrow = \omega_3 + \omega_m$,
the mechanical oscillator will experience the coupling to the third cavity as an effective negative temperature bath \cite{Marquardt2007PRL}. 
We therefore add $\hat{H}_\mathrm{aux} = \hbar \omega_3(\hat{x}) \hat{a}^\dagger_3 \hat{a}_3 + \im \hbar (\Omega_\uparrow \ex^{-i \omega_\uparrow t} \hat{a}^\dagger_3 - \mathrm{h.c.})$ to the Hamiltonian. We assume that the resonance frequency of the auxiliary mode is far away from the other mode frequencies, such that $|\omega_3 - \omega_{1,2}| \gg \omega_m$.

\emph{Mapping to the generic model}. 
The Hamiltonian can be made time independent \cite{SM} by going to a rotating frame at the drive frequency $\omega_\downarrow$ ($\omega_\uparrow$) for the modes $\hat{a}_\pm$ (mode $\hat{a}_3$), such that $\hat{a}_\pm \rightarrow \ex^{-\im \omega_\downarrow t} \hat{a}_\pm$ and $\hat{a}_3 \rightarrow \ex^{-\im \omega_\uparrow t} \hat{a}_3$. The two drives will create nonzero coherences in the modes $\hat{a}_-$ and $\hat{a}_3$. For this reason, we perform two displacement transformations (see details in \cite{SM}) $\hat{a}_- \rightarrow \bar{a}_- + \hat{a}_-$ and $\hat{a}_3 \rightarrow \bar{a}_3 + \hat{a}_3$, such that $\hat{a}_-$ and $\hat{a}_3$ now describe fluctuations around the mean cavity amplitudes $\bar{a}_- = \Omega_\downarrow/(\kappa_-/2 + \im \omega_m)$ and $\bar{a}_3 = \Omega_\uparrow/(\kappa_3/2 - \im \omega_m)$. We also define the detuning $\Delta_+ = \omega_\downarrow - \omega_+$.

The density matrix for the total system $\hat{\chi}_\mathrm{tot}$ is then determined by the quantum master equation $\dot{\hat{\chi}}_\mathrm{tot} = -(i/\hbar)[\hat{H}_\mathrm{tot} \, , \hat{\chi}_\mathrm{tot}] + {\cal L}_{d,\mathrm{tot}}\hat{\chi}_\mathrm{tot}$ with the total Hamiltonian
\begin{eqnarray}
\label{eq:Htotal}
\hat{H}_\mathrm{tot} & = & - \hbar \Delta_+ \hat{a}^\dagger_+ \hat{a}_+ + \hbar \omega_m \left(\hat{c}^\dagger \hat{c} + \hat{a}_-^\dagger \hat{a}_- - \hat{a}^\dagger_3 \hat{a}_3 \right) \\
& + & \hbar \left(\hat{c} + \hat{c}^\dagger \right) \Big[G_\downarrow \left(\hat{a}_+ + \hat{a}^\dagger_+\right) + G_\uparrow \left(\hat{a}_3 + \hat{a}^\dagger_3\right)  \notag \\
& + &    g_{++} \hat{a}^\dagger_+ \hat{a}_+ + g_{+-} \left(\hat{a}^\dagger_+ \hat{a}_- + \hat{a}^\dagger_- \hat{a}_+ \right) + g_3 \hat{a}^\dagger_3 \hat{a}_3 \Big] \notag \ .
\end{eqnarray}
where $G_\downarrow = g_{+-} \bar{a}_-$ and $G_\uparrow = g_3 \bar{a}_3$. We will define $G_\downarrow$ and $G_\uparrow$ to be real and positive, without loss of generality. We have neglected a term $\hbar g_3 (\hat{c} + \hat{c}^\dagger)|\bar{a}_3|^2$, which corresponds to redefining $x_0$ \cite{SM}. The dissipative terms are given by
\begin{equation}
\label{eq:LdissOM}
{\cal L}_{d,\mathrm{tot}} = \sum_{\mu = \pm} \kappa_\mu {\cal D}[\hat{a}_\mu]  + \kappa_3 {\cal D}[\hat{a}_3] + \gamma_{\downarrow, \therm}  {\cal D}[\hat{c}] + \gamma_{\uparrow, \therm} {\cal D}[\hat{c}^\dagger] 
\end{equation}
with $\gamma_{\uparrow, \therm} = \gamma n_\therm$ and $\gamma_{\downarrow, \therm} = \gamma (n_\therm + 1) $.
In principle, there could also be off-diagonal dissipative terms involving both modes $\hat{a}_\pm$, but we show in the Supplementary Material \cite{SM} that such terms are small. Furthermore, the important dissipation channel will be due to the intentionally increased decay of the mode $\hat{a}_-$, whereas the off-diagonal terms can maximally be of the size of the intrinsic dissipation rate.

We assume $|g_{+-}| \ll \kappa_-$ and $G_\uparrow \ll \kappa_3$, which means that the modes $\hat{a}_-$ and $\hat{a}_3$ decay fast and will be empty most of the time (in the displaced frame). This fact allows us to derive an effective master equation for the reduced density matrix $\hat{\chi}_{+,m}$ describing the modes $\hat{a}_+$ and $\hat{c}$ only. The derivation is based on a projection operator technique \cite{Breuer2007Book}, and extensive details can be found in the Supplementary Material \cite{SM}.

The effective master equation for the reduced density matrix $\hat{\chi}_{+,m}$ still contains the bilinear interaction terms proportional to $G_\downarrow$ as in Eq.~\eqref{eq:Htotal}. These terms give rise to normal modes which are linear combinations of photons and phonons \cite{SM,Borkje2013PRL,Lemonde2013PRL,Liu2013PRL}. We assume $\Delta_+ \sim -2\omega_m$ and $G_\downarrow/\omega_m \ll 1$, which means that the normal modes do not differ much from the original photon and phonon modes \cite{Borkje2013PRL}. We can describe the system in terms of these normal modes by applying a unitary transformation $\hat{U}$. To lowest order in $G_\downarrow/\omega_m \ll 1$, the transformation gives $\hat{U}^\dagger \hat{a}_+ \hat{U} = \hat{a}_+ - G_\downarrow(\hat{c} + \hat{c}^\dagger/3)/\omega_m $ and $\hat{U}^\dagger \hat{c} \, \hat{U} = \hat{c} + G_\downarrow(\hat{a}_+ - \hat{a}_+^\dagger/3)/\omega_m $. We choose the ideal detuning $\Delta_+ = -2\omega_m (1 - 5G_\downarrow^2/(3\omega_m^2) + G_\uparrow^2/(2\omega_m^2))$ \cite{SM} and move to rotating frames for both $\hat{c}$ and $\hat{a}_+$. In the Supplementary Material \cite{SM}, we show that the transformed density matrix $\hat{\rho} = \hat{U}^\dagger \hat{\chi}_{+,m} \hat{U}$ is determined by the master equation $\dot{\hat{\rho}} = ({\cal L}_c + {\cal L}_d )\hat{\rho}$, defined by Eqs.~\eqref{eq:Lc} and \eqref{eq:Ld} when renaming $\hat{a}_+ \rightarrow \hat{a}$. We reiterate that the operators $\hat{a}$ and $\hat{c}$ now refer to normal modes which are almost, but not quite, the same as the original photon and phonon modes. The rates in ${\cal L}_c$ and ${\cal L}_d$ become $\tilde{g} = -g_{++}G_\downarrow/\omega_m$, $\Gamma = 4 g_{+-}^2/\kappa_-$, $\kappa = \kappa_+$, $\gamma_\downarrow = \gamma (n_\mathrm{th} + 1) + G^2_\uparrow \kappa_3/(2\omega_m)^2 + G_\downarrow^2 \kappa_+/\omega_m^2 $, and $\gamma_\uparrow = \gamma n_\mathrm{th} + 4 G_\uparrow^2/\kappa_3 + G^2_\downarrow \kappa_+/(3\omega_m)^2 $.
The last two terms in $\gamma_\downarrow \, (\gamma_\uparrow)$
originate from anti-Stokes (Stokes) scattering of photons from the two drives. 

In the desired regime, we can find accurate analytical expressions for the steady-state density matrix by truncating the Hilbert space \cite{SM}. After solving the steady state equation $\dot{\hat{\rho}} = 0$, we have to transform back to the original density matrix $\hat{\chi}_{+,m}$ in the basis of photons and phonons, but this only gives small corrections of order $(G_\downarrow/\omega_m)^2$ compared to the occupation probabilites given by $\hat{\rho}$ \cite{SM}.   

There are three requirements for the mechanical oscillator to settle into an $n = 1$ Fock state, as was discussed above. First of all, we need $\kappa \ll \Gamma$, which is satisfied when $4 g_{+-}^2/(\kappa_- \kappa_+) \gg 1$. Furthermore, $\gamma_\downarrow \ll \gamma_\uparrow$ follows when assuming $4 G^2_\uparrow/\kappa_3 \gg (G_\downarrow/\omega_m)^2 \kappa_+, \gamma (n_\therm + 1)$, which can in principle be achieved by increasing the drive strength $|\Omega_\uparrow|$. Finally, for the ratio $P_2/P_1$ to be small, we must require $\gamma_\uparrow \ll \Gamma$ as well as $\gamma_\uparrow \Gamma/(4 \tilde{g}^2) \ll 1$, which puts an upper limit on $|\Omega_\uparrow|$. Note that the criteria requires $\gamma_\uparrow \Gamma/(4 \tilde{g}^2) \gg \gamma_\downarrow \Gamma/(4 \tilde{g}^2) \geq (g_{+-}/g_{++})^2 \kappa_+/\kappa_-$ which limits how close one can get to a pure $n=1$ phonon Fock state in this particular realization. We note that our scheme is not explicitly dependent on the size of the mechanical frequency $\omega_m$, since the ratio $G_\downarrow/\omega_m$ is controlled by the drive power. This is in contrast to the nonlinear effects discussed in Refs.~\cite{Rabl2011PRL,Nunnenkamp2011PRL}. We also note that if $g_{--} \neq 0$ due to imperfections, the scheme still works as long as $4|g_{--} \bar{a}_-|^2/\kappa_- \ll \gamma_\uparrow$. 
 
\begin{figure}[h]
 \centering
 \includegraphics[width=0.99\columnwidth,trim=0cm 0cm 0cm 0cm]{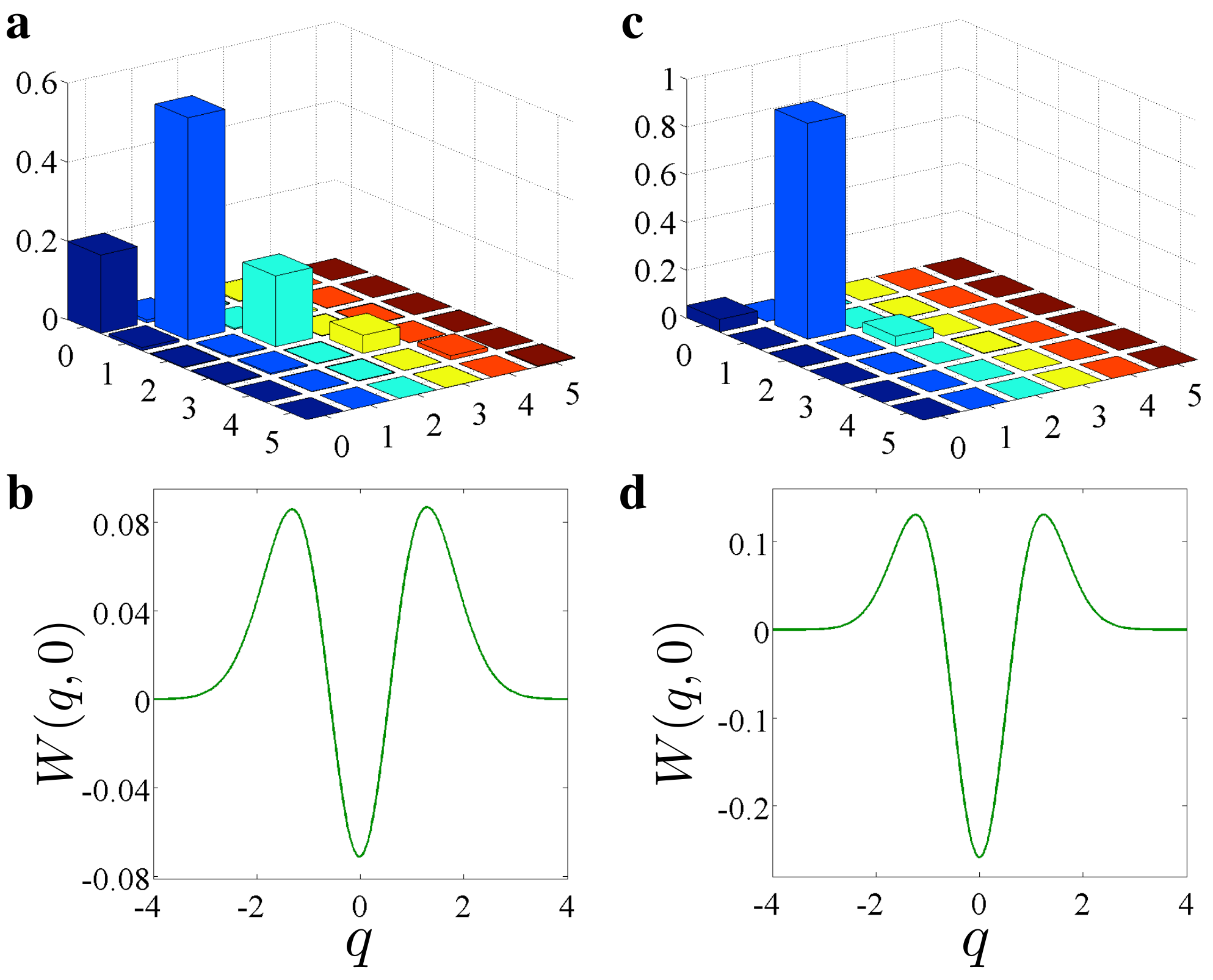}
\caption{(color online). Numerical results. (a,c): Reduced steady-state density matrix $\hat{\chi}_m$ for the mechanical oscillator in the Fock basis. (b,d): The Wigner distribution $W(q,p)$ for $p = 0$. $W(q,p)$ is rotationally symmetric around $q=p=0$ \cite{SM}. Parameters for (a,b) are $g_{++}/\kappa_+ = g_{+-}/\kappa_+ = 10$, $g_3/\kappa_+ = 0.1$, $\omega_m/\kappa_+ = 200$, $\kappa_-/\kappa_+ = \kappa_3/\kappa_+ = 50$, $G_\downarrow/\omega_m = 0.1$, $G_\uparrow/\kappa_3 = 0.026$, $\omega_m/\gamma = 10^6$, and $n_\mathrm{th} = 10$. For (c,d), we used $g_{++}/\kappa_+ = g_{+-}/\kappa_+ = 200$, $g_3/\kappa_+ = 0.1$, $\omega_m/\kappa_+ = 5 \times 10^3$, $\kappa_-/\kappa_+ = \kappa_3/\kappa_+ = 10^3$, $G_\downarrow/\omega_m = 0.05$, $G_\uparrow/\kappa_3 = 5.3 \times 10^{-3}$, $\omega_m/\gamma = 10^7$, and $n_\mathrm{th} = 1$. In both cases, we see that the oscillator is in a nonclassical steady state, indicated by negativity of the Wigner distribution in the central region of phase space.}
\label{fig:Plots}
\end{figure}

\emph{Negative Wigner distribution}.
Tracing over the three optical cavity modes gives the reduced density matrix for the mechanical oscillator $\hat{\chi}_m = \mathrm{Tr}_\mathrm{opt} \hat{\chi}_\mathrm{tot}$. This can be represented by its associated Wigner distribution $W(q,p)$ \cite{SM}, which in the classical limit can be interpreted as a phase space probability distribution. 
Even if the parameters are not ideal for preparing a pure Fock state, the scheme presented here can stabilize the mechanical oscillator in a steady state which is nonclassical in the sense that the Wigner distribution has regions of negativity. This is demonstrated in Fig.~\ref{fig:Plots}, which shows the result of solving the quantum master equation defined by Eqs.~\eqref{eq:Htotal} and \eqref{eq:LdissOM} numerically. The parameters used in Figs.~3(a-b) might be within reach of experiments. The ones used in Figs.~3(c-d) are not very realistic, but show that the model produces an almost pure Fock state in the ideal regime, as the $n = 1$ occupation probability exceeds 0.9.

\emph{Detection}.
The steady-state Wigner distribution can in principle be obtained from optomechanical back-action free quadrature detection \cite{Braginsky1995Book,Clerk2008NJP} and subsequent quantum state tomography \cite{Lvovsky2009RMP}. Additionally, sideband thermometry \cite{Marquardt2007PRL} 
can be used to measure the ratio $\bar{n}/(\bar{n} + 1)$, where $\bar{n} = \langle c^\dagger c \rangle$, which would asymptotically approach 1/2 as the oscillator state approaches the Fock state.

\emph{Conclusion}.
We have studied a generic reservoir engineering scheme that autonomously stabilizes a harmonic oscillator in an $n=1$ Fock state. As a physical example, we analyzed an optomechanical setup where a mechanical oscillator is strongly coupled to several optical cavity modes. We showed, both analytically and numerically, that the mechanical oscillator relaxes into a nonclassical state in a certain parameter regime, and that this state approaches the $n=1$ Fock state in the ideal limit.
 
\emph{Acknowledgements.}
The author acknowledges financial support from The Danish Council for Independent Research under the Sapere Aude program, as well as useful input from Max Ludwig, Florian Marquardt, Steve Girvin, and Andreas Nunnenkamp. The numerical calculations were performed with the Quantum Optics Toolbox \cite{Tan1999}.

%



\newpage
\begin{center}
{\large {\bf Supplementary Material to ``Scheme for steady-state preparation of a harmonic oscillator in the first excited state''}}\\
\end{center}

\section{1. Approximate analytical solution to generic model}

We now derive an approximate solution to the generic quantum master equation $\dot{\hat{\rho}} = \left({\cal L}_c + {\cal L}_d\right) \hat{\rho}$ in the regime $\kappa \ll \Gamma$, $\gamma_\downarrow \ll  \gamma_\uparrow $, $\gamma_\uparrow \ll \mathrm{min}(4 \tilde{g}^2/\Gamma, \Gamma)$. Let a general state in the Fock basis be written as $|k\rangle \langle k'| \otimes |n\rangle \langle n'|$ where $k,k'$ refers to system $\hat{a}$ and $n,n'$ to the oscillator $\hat{c}$. In case system $\hat{a}$ is a two-level system ($\hat{a} \rightarrow \hat{\sigma}_-$), $k = 0$ refers to the ground state and $k =1 $ to the excited state. We make the ansatz
\begin{eqnarray}
\label{eq:ansatzrho}
\hat{\rho} & = & |0 \rangle \langle 0| \otimes \left(\rho_{00,0} |0\rangle \langle 0| + \rho_{00,1} |1\rangle \langle 1| \right. \\
& & \qquad \quad +  \left. \rho_{00,2} |2\rangle \langle 2| + \rho_{00,3} |3\rangle \langle 3|\right) \nonumber\\ 
& + & |1 \rangle \langle 0| \otimes \left(\rho_{10,0} |0\rangle \langle 2| + \rho_{10,1} |1\rangle \langle 3|\right) \nonumber \\
& + & |0 \rangle \langle 1| \otimes \left(\rho_{10,0}^\ast |2\rangle \langle 0| + \rho_{10,1}^\ast |3\rangle \langle 1|\right) \nonumber \\ 
& + & |1 \rangle \langle 1| \otimes \left(\rho_{11,0} |0\rangle \langle 0| + \rho_{11,1} |1\rangle \langle 1|\right) \nonumber .
\end{eqnarray}
By truncating the Hilbert space to $n,n' < 4$, we can determine the coefficients $\rho_{mm',n}$ defined in \eqref{eq:ansatzrho} by inserting the ansatz into the master equation. To lowest order in $\kappa/\Gamma$, $\gamma_\uparrow/\Gamma$, $\gamma_\downarrow/\Gamma$, and $\gamma_\uparrow \Gamma/(4 \tilde{g}^2)$, we find 
\begin{eqnarray}
\label{eq:SolutionGeneric}
\frac{\rho_{00,0}}{\rho_{00,1}} & = & \frac{\gamma_\downarrow}{\gamma_\uparrow} + \frac{2\kappa}{\Gamma}   \\
\frac{\rho_{00,2}}{\rho_{00,1}} & = &  \frac{\gamma_\uparrow \Gamma}{4 \tilde{g}^2} +  \frac{2\gamma_\uparrow}{\Gamma}   \notag \\
\frac{\rho_{00,3}}{\rho_{00,1}} & = & \left( \frac{\gamma_\uparrow \Gamma}{4 \tilde{g}^2} \right)^2 \left[ 1 + 6 \left(\frac{\tilde{g}}{\Gamma}\right)^2 + 64 \left(\frac{\tilde{g}}{\Gamma}\right)^4 \right]  \notag \\
\frac{\rho_{11,0}}{\rho_{00,1}} & = & \frac{2\gamma_\uparrow}{\Gamma}  \notag \\
\frac{\rho_{11,1}}{\rho_{00,2}} & = & \frac{\gamma_\uparrow}{\Gamma} \left(\frac{3}{2} + \frac{4 (\tilde{g}/\Gamma)^2}{1 + 8 (\tilde{g}/\Gamma)^2} \right)  \notag \\
\frac{\rho_{10,0}}{\rho_{00,2}} & = & \frac{-2 \sqrt{2} i \tilde{g}/\Gamma}{1 + 8 (\tilde{g}/\Gamma)^2}  \notag \\
\frac{\rho_{10,1}}{\rho_{00,2}} & = & \frac{-6 i \tilde{g}/\Gamma}{1 + 6 (\tilde{g}/\Gamma)^2} \left[ \frac{\rho_{00,3}/\rho_{00,1}}{\rho_{00,2}/\rho_{00,1}} + \frac{2 \gamma_\uparrow (1 - 2 (\tilde{g}/\Gamma)^2)}{\Gamma(1 + 8 (\tilde{g}/\Gamma)^2)}  \right]. \notag
\end{eqnarray}
This leaves one unknown, $\rho_{00,1}$, which is straightforwardly determined from the criterion $\mathrm{Tr} \, \hat{\rho} = \sum_{m,n} \rho_{mm,n} = 1$, giving
\begin{equation}
\label{eq:solrho001}
\rho_{00,1} = \frac{1}{1 + \frac{\rho_{00,0}}{\rho_{00,1}} + \frac{\rho_{00,2}}{\rho_{00,1}} + \frac{\rho_{00,3}}{\rho_{00,1}} + \frac{\rho_{11,0}}{\rho_{00,1}}  + \frac{\rho_{11,1}}{\rho_{00,2}}  \frac{\rho_{00,2}}{\rho_{00,1}}} .
\end{equation} 

With the ansatz \eqref{eq:ansatzrho}, the probabilities $P_n$ to find oscillator $\hat{c}$ in the Fock state $|n\rangle$ become
\begin{eqnarray}
\label{eq:Pn}
P_0 & = & \rho_{00,0} + \rho_{11,0} \\
P_1 & = & \rho_{00,1} + \rho_{11,1} \notag \\
P_2 & = & \rho_{00,2} \notag \\
P_3 & = & \rho_{00,3} \notag 
\end{eqnarray}
and $P_n = 0$, $n > 3$. We have compared these analytical results to numerical calculations of $\rho$ from the generic model and found very good agreement. Also, in Sec.~5, we use these results to compare with numerical calculations on the optomechanical model.

\section{2. Effective cavity modes and inter- and intramode optomechanical coupling}

The optical part of the Hamiltonian $\hat{H}_\mathrm{free}$ is
\begin{equation}
\label{eq:Hfreeopt}
\hat{H}_\mathrm{free, opt} = \hbar \omega_1 \hat{a}_1^\dagger \hat{a}_1 + \hbar \omega_2 \hat{a}_2^\dagger \hat{a}_2 + \hbar J \left(\hat{a}_1^\dagger \hat{a}_2 + \hat{a}_2^\dagger \hat{a}_1 \right),
\end{equation}
which is diagonalized by the transformation 
\begin{eqnarray}
\label{eq:Transf}
\hat{a}_+ & = & \frac{1}{\sqrt{1 + r^2}} \left(r \hat{a}_1 + \hat{a}_2 \right) , \\
\hat{a}_- & = & \frac{1}{\sqrt{1 + r^2}} \left(\hat{a}_1 - r \hat{a}_2 \right) , \notag
\end{eqnarray}
with 
\begin{eqnarray}
\label{eq:rdef}
r & = & \frac{2J/\delta}{1 + \mathrm{sgn}(\delta) \sqrt{1 + (2J/\delta)^2}}, \\
\delta & = & \omega_2 - \omega_1 ,
\end{eqnarray}
giving $\hat{H}_\mathrm{free, opt} = \hbar \sum_{\mu = \pm} \omega_\mu \hat{a}^\dagger_\mu \hat{a}_\mu$ with the frequencies
\begin{equation}
\label{eq:Freqs}
\omega_\pm = \frac{\omega_1 + \omega_2 \pm \sqrt{(\omega_2 - \omega_1)^2 + 4J^2}}{2}.
\end{equation}

The reverse transform is
\begin{eqnarray}
\label{eq:RevTransf}
\hat{a}_1 & = & \frac{1}{\sqrt{1 + r^2}} \left(\hat{a}_- + r \hat{a}_+ \right) , \\
\hat{a}_2 & = & \frac{1}{\sqrt{1 + r^2}} \left(\hat{a}_+ - r \hat{a}_- \right) , \notag
\end{eqnarray}
which can be used to express the optomechanical interaction Hamiltonian $\hat{H}_\mathrm{int}$ in terms of the effective cavity modes. This gives $\hat{H}_\mathrm{int} = \hbar \left(\hat{c} + \hat{c}^\dagger \right) \sum_{\mu,\nu = \pm} g_{\mu \nu} \hat{a}^\dagger_\mu \hat{a}_\nu$ with
\begin{eqnarray}
\label{eq:geff}
g_{++} & = & \frac{r^2 g_1  + g_2}{1 + r^2}, \\
g_{--} & = & \frac{g_1 +  r^2 g_2}{1 + r^2}, \\
g_{+-} & = & \frac{r(g_1 - g_2)}{1 + r^2},
\end{eqnarray}
and $g_{-+} = g_{+-}$. To make the intraband coupling $g_{--} = 0$, we must assume $\mathrm{sgn}(g_1 g_2) = -1$ and the ratio $|J/\delta|$ must be chosen such that
\begin{equation}
\label{eq:rchoice}
|r| = \sqrt{\left|\frac{g_1}{g_2}\right|} ,
\end{equation}
which means that $g_1 = -r^2 g_2$. With that choice of $r$, the other coupling rates become
\begin{equation}
\label{eq:geffChoice}
g_{++} = (1 - r^2)g_2 , \quad g_{+-} = -r g_2 .
\end{equation}
We see that the scheme does not work if the coupling rates $|g_1| = |g_2|$, since that requires $|r| = 1$ which gives $g_{++} = 0$. Also, we cannot start with degenerate modes ($\delta = 0$), since that also gives $|r| = 1$.

We want both coupling rates $|g_{++}|$ and $|g_{+-}|$ to be comparable to the orginal rates $|g_1|, |g_2|$. This is achieved for a wide range of values $|r|$. As a special case, let us examine at which $|r| = |r_e|$ the rates $|g_{++}|$ and $|g_{+-}|$ are equal. This requires  
\begin{equation}
\label{eq:equalgeffs}
 |r_e| = |1 - r_e^2| .
\end{equation}
This equation is in fact the one determining the golden ratio and has two solutions that are each others inverse, $|r_e| = 0.62$ and $|r_e| = 1.62$. Curiously, this particular case is realized when the tunneling rate $|J|$ equals the detuning $|\delta|$.

\section{3. Derivation of the quantum master equation for the optomechanical system}

\subsection{3.1. Cavity mode dissipation}

\subsubsection{3.1.1. Intrinsic}

We now examine the dissipation experienced by the effective cavity modes $\hat{a}_+$ and $\hat{a}_-$. This can be done at the master equation level (see e.g.~Ref.~\cite{Carmichael1973JPhysA}), but we will here use an approach based on quantum Langevin equations. \cite{Clerk2010RMP}

The physical cavity modes $\hat{a}_1$ and $\hat{a}_2$ are coupled to uncontrolled degrees of freedom in the environment. This can usually be modeled by coupling to a bath of harmonic oscillators, with a system-bath coupling
\begin{equation}
\label{eq:Hsb}
\hat{H}_\mathrm{s-b} = i \hbar \sum_{j=1,2} \sum_k \lambda_{j,k} \left(\hat{f}_{j,k}^\dagger \hat{a}_j - \hat{a}_j^\dagger \hat{f}_{j,k} \right)
\end{equation}
Here, $k$ is a quantum number (or set of quantum numbers) numerating the bath modes. We assume for simplicity that the bath modes form a discrete set, but will take the continuum limit later. The real number $\lambda_{j,k}$ characterizes the coupling strength between cavity mode $j$ and bath mode $\hat{f}_{j,k}$. The bath mode Hamiltonian is
\begin{equation}
\label{eq:Hb}
\hat{H}_\mathrm{b} = \hbar \sum_{j=1,2} \sum_k \nu_{j,k} \hat{f}_{j,k}^\dagger \hat{f}_{j,k} ,
\end{equation}
where $\nu_{j,k} $ are the bath mode frequencies, and the system Hamiltonian is $\hat{H}_\mathrm{s} = \hat{H}_\mathrm{free, opt}$ given in Eq.~\eqref{eq:Hfreeopt}. We can ignore the coupling to the mechanical oscillator here.

By inserting the reverse transform \eqref{eq:RevTransf} in the Hamiltonian \eqref{eq:Hsb}, we can now derive the Heisenberg equations for the effective cavity modes $\hat{a}_+$ and $\hat{a}_-$ and for the bath modes: 
\begin{eqnarray}
\label{eq:Heisenberg}
\dot{\hat{a}}_j & = & \frac{i}{\hbar} \left[\hat{H}_\mathrm{s} + \hat{H}_\mathrm{s-b} + \hat{H}_\mathrm{b} \, , \, \hat{a}_j\right] \\
 \dot{\hat{f}}_{j,k} & = & \frac{i}{\hbar} \left[\hat{H}_\mathrm{s} + \hat{H}_\mathrm{s-b} + \hat{H}_\mathrm{b} \, , \, \hat{f}_{j,k} \right] .
\end{eqnarray}
These equations will be coupled, but we can eliminate the bath operators to find
\begin{eqnarray}
\label{eq:Langevin1}
& & \dot{\hat{a}}_+ =  -i \omega_+ \hat{a}_+ \\
& & \ - \frac{r}{\sqrt{1+r^2}} \sum_k \lambda_{1,k} \ex^{-i \nu_{1,k} (t-t_0)} \hat{f}_{1,k}(t_0) \notag \\
& & \ -  \frac{r}{1+r^2}  \sum_k\lambda_{1,k}^2 \int_{t_0}^t d \tau \, \ex^{-i \nu_{1,k} (t-\tau)} \left(\hat{a}_-(\tau) + r\hat{a}_+(\tau)\right) \Big] \notag \\
& & \ - \frac{1}{\sqrt{1+r^2}} \sum_k \lambda_{2,k} \ex^{-i \nu_{2,k} (t-t_0)} \hat{f}_{2,k}(t_0) \notag \\
& & \ -  \frac{1}{1+r^2}  \sum_k\lambda_{2,k}^2 \int_{t_0}^t d \tau \, \ex^{-i \nu_{2,k} (t-\tau)} \left(\hat{a}_+(\tau) - r\hat{a}_-(\tau)\right) \Big] \notag  
\end{eqnarray}
and
\begin{eqnarray}
\label{eq:Langevin2}
& & \dot{\hat{a}}_- =  -i \omega_- \hat{a}_- \\
& & \ - \frac{1}{\sqrt{1+r^2}} \sum_k \lambda_{1,k} \ex^{-i \nu_{1,k} (t-t_0)} \hat{f}_{1,k}(t_0) \notag \\
& & \ -  \frac{1}{1+r^2}  \sum_k\lambda_{1,k}^2 \int_{t_0}^t d \tau \, \ex^{-i \nu_{1,k} (t-\tau)} \left(\hat{a}_-(\tau) + r\hat{a}_+(\tau)\right) \Big] \notag \\
& & \ + \frac{r}{\sqrt{1+r^2}} \sum_k \lambda_{2,k} \ex^{-i \nu_{2,k} (t-t_0)} \hat{f}_{2,k}(t_0) \notag \\
& & \ +  \frac{r}{1+r^2}  \sum_k\lambda_{2,k}^2 \int_{t_0}^t d \tau \, \ex^{-i \nu_{2,k} (t-\tau)} \left(\hat{a}_+(\tau) - r\hat{a}_-(\tau)\right) \Big] \notag , 
\end{eqnarray}
where we let $t_0$ be a time in the distant past. We now take the continuum limit, such that
\begin{eqnarray}
\label{eq:continuum}
S_j(t - \tau) & \equiv & \sum_k \lambda^2_{j,k} \ex^{-i \nu_{j,k} (t-\tau)} \\
& \rightarrow & \int d \omega \, D_j(\omega) \tilde{\lambda}^2_j(\omega) \ex^{-i \omega (t-\tau)} , \notag
\end{eqnarray}
where $D_j(\omega)$ is a density of states and $\tilde{\lambda}_j(\omega) = \lambda_{k(\omega),j}$. 
 
The density of states $D_j(\omega)$ and the coupling strenghts $\tilde{\lambda}_j(\omega)$ can be complicated functions of frequency. However, we are only interested in their values in a narrow frequency range arond the cavity resonance frequencies whose width is on the order of $\omega_m$. We will therefore treat them as constants, i.e.~$D_j(\omega) \approx D_j$ and $\tilde{\lambda}_j(\omega) \approx \tilde{\lambda}_j$, which is likely to be a very good approximation. The sums $S_j$ then become
\begin{equation}
\label{eq:Sj}
S_j(t - \tau) \approx 2\pi D_j \tilde{\lambda}_j \delta(t- \tau)
\end{equation}
such that the $\tau$-integrals become trivial. 

The quantum Langevin equations then become
\begin{eqnarray}
\label{eq:qLe}
\dot{\hat{a}}_+ & = & - \left(\frac{\kappa_{+}}{2} + i \omega_+\right) \hat{a}_+ - \frac{\kappa_{+-}}{2} \hat{a}_- + \sqrt{\kappa_+}\hat{\xi}_+ \\ 
\dot{\hat{a}}_- & = & - \left(\frac{\kappa_{-}}{2} + i \omega_-\right) \hat{a}_- - \frac{\kappa_{+-}}{2} \hat{a}_+ + \sqrt{\kappa_-}\hat{\xi}_- \notag
\end{eqnarray}
when we define the parameters
\begin{eqnarray}
\label{eq:kappaDef}
\kappa_+ & = & \frac{r^2 \kappa_1 + \kappa_2}{1 + r^2} \\
\kappa_- & = & \frac{\kappa_1 + r^2 \kappa_2}{1 + r^2} \notag \\
\kappa_{+-} & = & \frac{r}{1 + r^2}(\kappa_1 - \kappa_2) \notag 
\end{eqnarray}
with
\begin{equation}
\label{eq:kappa12Def}
\kappa_j = 2\pi \tilde{\lambda}_j^2 D_j , \quad j = 1,2,
\end{equation}
and the vacuum noise operators
\begin{eqnarray}
\label{eq:NoiseDef}
\hat{\xi}_+(t) & = & \frac{1}{\sqrt{1+r^2}} \left(r \sqrt{\frac{\kappa_1}{\kappa_+}} \hat{\xi}_1(t) + \sqrt{\frac{\kappa_2}{\kappa_+}} \hat{\xi}_2(t) \right) \\
\hat{\xi}_-(t) & = & \frac{1}{\sqrt{1+r^2}} \left(\sqrt{\frac{\kappa_1}{\kappa_-}} \hat{\xi}_1(t) - r \sqrt{\frac{\kappa_2}{\kappa_-}} \hat{\xi} _2(t) \right) \notag
\end{eqnarray}
with
\begin{equation}
\label{eq:Noise12Def}
\hat{\xi}_j(t) = - \frac{1}{\sqrt{2\pi D_j}} \sum_k \ex^{-i \nu_{j,k} (t - t_0)} \hat{f}_{j,k}(t_0), \quad j = 1,2.
\end{equation}
The properties of the effective vacuum noise operators are
\begin{eqnarray}
\label{eq:CorrNoise}
\langle \hat{\xi}_+(t) \hat{\xi}^\dagger_+(t') \rangle & = & \delta(t-t') \\ 
\langle \hat{\xi}_-(t) \hat{\xi}^\dagger_-(t') \rangle & = & \delta(t-t') \notag \\
\langle \hat{\xi}_+(t) \hat{\xi}^\dagger_-(t') \rangle & = & \frac{\kappa_{+-}}{\sqrt{\kappa_+ \kappa_-}} \delta(t-t') , \notag
\end{eqnarray}
which follow from
\begin{equation}
\label{eq:CorrNoise12}
\langle \hat{\xi}_j(t) \hat{\xi}^\dagger_j(t') \rangle = \delta(t - t') , \quad j = 1,2,
\end{equation}
which again follow from assuming that the bath modes are in the vacuum state.

Note that Eqs.~\eqref{eq:qLe} is exactly what we would have if we included dissipation for the cavities $\hat{a}_1$ and $\hat{a}_2$ separately before taking into account the tunneling $J$, i.e.~if we had started from the equations
\begin{eqnarray}
\label{eq:LangevinPhysical}
\dot{\hat{a}}_1 & = & - \left(\frac{\kappa_{1}}{2} + i \omega_1\right) \hat{a}_1 - i J \hat{a}_2 + \sqrt{\kappa_1}\hat{\xi}_1 \\
\dot{\hat{a}}_2 & = & - \left(\frac{\kappa_{2}}{2} + i \omega_2\right) \hat{a}_2 - i J \hat{a}_1 + \sqrt{\kappa_2}\hat{\xi}_2 . \notag
\end{eqnarray}
We emphasize that this is in general the wrong approach unless the modes only hybridize very weakly, i.e.~if $|J| \ll \mathrm{max}(|\delta|,\kappa)$, which is not the case here. The reason why it nevertheless works here is that we made the assumptions of constant $D_j$ and $\tilde{\lambda}_j$ in the relevant frequency regime.

This means that, when returning to the master equation, the terms describing dissipation of the effective modes $\hat{a}_+$ and $\hat{a}_-$ are given by
\begin{eqnarray}
\label{eq:DissTerms}
{\cal L}_{d,+-} \hat{X} & = & \kappa_1 {\cal D}[\hat{a}_1] \hat{X} + \kappa_2 {\cal D}[\hat{a}_2] \hat{X} \\
& = & \kappa_+ {\cal D}[\hat{a}_+] \hat{X} + \kappa_- {\cal D}[\hat{a}_-]\hat{X} \notag \\ & + & \kappa_{+-} \Big(\hat{a}_- \hat{X} \hat{a}_+^\dagger + \hat{a}_+ \hat{X} \hat{a}_-^\dagger \notag \\
& & \qquad - \frac{1}{2}\left\{\hat{a}_+^\dagger \hat{a}_- + \hat{a}_-^\dagger \hat{a}_+ \, , \, \hat{X} \right\} \Big) \notag .
\end{eqnarray}
We see that there are cross-terms proportional to $\kappa_{+-}$ that we have not included in the model in the main article. The omission of these terms can certainly be justified in the special case when the intrinsic dissipation of the physical cavities are the same, i.e.~when $\kappa_1 = \kappa_2$, since $\kappa_{+-} = 0$ in that case. A more general justification for omitting them is the assumption that $\kappa_-$ is enhanced by external coupling to the effective mode $\hat{a}_-$ (see next section), such that $\kappa_- \gg \kappa_+, \kappa_{+-}$. This means that the dissipative terms $\kappa_+$ and $\kappa_{+-}$ will not play a major role as long as $\kappa_+, \kappa_{+-} \ll \Gamma$, which is a necessary assumption for the scheme to work anyway. Thus, in the regime we are interested in, the terms proportional to $\kappa_{+-}$ play no significant role and we may neglect them.

\subsubsection{3.1.2. Extrinsic}

The assumptions that $g_{--} = 0$ and that $g_{++}, g_{+-}$ are on the order of the original couplings $g_1,g_2$ means that $r \sim {\cal O}(1)$. This again means that the modes $\hat{a}_1$ and $\hat{a}_2$ strongly hybridize and that $\kappa_+$ and $\kappa_-$, as defined in Eq.~\eqref{eq:kappaDef}, are of the same order of magnitude. To achieve $\kappa_- \gg \kappa_-$, we therefore need to increase $\kappa_-$ by external coupling to the effective mode $\hat{a}_-$ only. To model this, let us assume that a third optical bath couples to the physical cavities $\hat{a}_1$ and $\hat{a}_2$ according to the Hamiltonian
\begin{equation}
\label{eq:Hsbext}
\hat{H}_\mathrm{s-b,ext} = i \hbar \sum_{j=1,2} \lambda_{j,\mathrm{ext}} \sum_k  \left(\hat{f}_{j,k,\mathrm{ext}}^\dagger \hat{a}_j - \hat{a}_j^\dagger \hat{f}_{j,k,\mathrm{ext}} \right) . 
\end{equation}
This is similar to Eq.~\eqref{eq:Hsb}, but we have already assumed that the coupling strength is constant for the bath modes in the frequency range that contributes. For this third bath to effectively couple to the mode $\hat{a}_-$ only, we need to assume that the couplings obey
\begin{equation}
\label{eq:lambdaext}
\frac{\lambda_{2,\mathrm{ext}}}{\lambda_{1,\mathrm{ext}}} = -r .
\end{equation}
According to Eqs.~\eqref{eq:rdef} and \eqref{eq:rchoice}, this means that
\begin{equation}
\label{eq:lambdaAbs}
\left|\frac{\lambda_{2,\mathrm{ext}}}{\lambda_{1,\mathrm{ext}}}\right| = \sqrt{\left|\frac{g_1}{g_2}\right|} 
\end{equation}
and that
\begin{equation}
\label{eq:lambdaSign}
 \mathrm{sgn}\left(\lambda_{1,\mathrm{ext}}\lambda_{2,\mathrm{ext}} J\right) = -1 . 
\end{equation}
Eq.~\eqref{eq:lambdaAbs} simply states how strongly the physical cavities should couple to the third bath, whereas Eq.~\eqref{eq:lambdaSign} is a requirement on the relative sign of the three couplings between cavity 1, cavity 2 and the external bath. If for example the cavity-bath couplings are both positive, we must have $J < 0$. Oppositely, if $J > 0$, the cavity-bath couplings must have opposite signs.  

\subsection{3.2. Rotating frame and displacement transformations}
Let us denote the density matrix of the total optomechanical system as $\tilde{\chi}_\mathrm{tot}$. The quantum master equation is
\begin{equation}
\label{eq:FullQMEPre}
\dot{\tilde{\chi}}_\mathrm{tot} =  -\frac{i}{\hbar} \left[\tilde{H}, \tilde{\chi}_\mathrm{tot} \right] + {\cal L}_{d,\mathrm{tot}}\tilde{\chi}_\mathrm{tot} \notag 
\end{equation}
with the Hamiltonian
\begin{equation}
\label{eq:tildeH}
\tilde{H} = \hat{H}_\mathrm{free} + \hat{H}_\mathrm{int} + \hat{H}_\mathrm{drive} + \hat{H}_\mathrm{aux} .
\end{equation}
To move to a frame where the Hamiltonian is time-independent, we perform a unitary transformation 
\begin{equation}
\label{eq:chiTilde}
\bar{\chi}_\mathrm{tot} = U_r^\dagger \tilde{\chi}_\mathrm{tot} U_r ,
\end{equation}
with 
\begin{equation}
\label{eq:Urot}
U_r = \ex^{-i \left[\omega_\downarrow (\hat{a}_+^\dagger \hat{a}_+ + \hat{a}_-^\dagger \hat{a}_-) + \omega_\uparrow \hat{a}_3^\dagger \hat{a}_3 \right] t} .
\end{equation}
The transformed density matrix $\bar{\chi}_\mathrm{tot}$ obeys the master equation
\begin{equation}
\label{eq:FullQMEPreRot}
\dot{\bar{\chi}}_\mathrm{tot} = -\frac{i}{\hbar} \left[\bar{H} , \bar{\chi}_\mathrm{tot} \right] + {\cal L}_{d,\mathrm{tot}}\bar{\chi}_\mathrm{tot} \notag ,
\end{equation}
with the Hamiltonian
\begin{equation}
\label{eq:HamiltonTimeIndep}
\bar{H} = U_r^\dagger \tilde{H} U_r - i U_r^\dagger \partial_t U_r ,
\end{equation}
which is time-independent.

We now perform the displacement transformations by defining
\begin{equation}
\label{eq:chitot}
\hat{\chi}_\mathrm{tot} = U_d^\dagger \bar{\chi}_\mathrm{tot} U_d ,
\end{equation}
with 
\begin{equation}
\label{eq:DisTransfDef}
U_d = \ex^{\bar{a}_- \hat{a}_-^\dagger - \bar{a}_-^\ast \hat{a}_- + \bar{a}_3 \hat{a}_3^\dagger - \bar{a}_3^\ast \hat{a}_3} .
\end{equation}
This gives the quantum master equation presented in the article, except for an additional term $\hbar g_3 (\hat{c} + \hat{c}^\dagger)|\bar{a}_3|^2$ in the Hamiltonian. This term will produce a nonzero expectation value of $\hat{c} + \hat{c}^\dagger$, which contradicts our assumption $\langle \hat{x} \rangle = x_0$. The error stems from the fact that we defined $x_0$ without taking into account the average displacement of the oscillator due to the drive $\Omega_\uparrow$. We could have started with a different definition of $x_0$ that took this into account, but it would simply have shifted the resonance frequencies $\omega_j$, $j = 1,2,3$ and not changed the physical picture at all. We may therefore assume that these shifts have already been included in the resonance frequencies and neglect this addition to the Hamiltonian.

\section{4. Mapping of the optomechanical system to the generic model}

\subsection{4.1. Projection operator technique}

In the limits $|g_{+-}| \ll \kappa_-$ and $G_\uparrow \ll \kappa_3$, the modes $\hat{a}_-$ and $\hat{a}_3$ are almost in the vacuum state (after the displacement transformations) and can be projected out. To do this, we follow Ref.~\cite{Breuer2007Book_SM} and define
\begin{eqnarray}
\label{eq:L1def}
\hat{H}^{(1)} & = &\hbar \left(\hat{c} + \hat{c}^\dagger \right) \Big\{g_{+-}\left(\hat{a}_+^\dagger \hat{a}_- + \hat{a}_-^\dagger \hat{a}_+ \right) \\
 & & \qquad  \qquad  \quad G_\uparrow \left(\hat{a}_3 + \hat{a}_3^\dagger \right) \Big\} , \notag \\ 
{\cal L}^{(1)} \hat{\chi}_\mathrm{tot} & = & -\frac{i}{\hbar} \left[\hat{H}^{(1)} , \hat{\chi}_\mathrm{tot} \right] , 
\end{eqnarray}
and
\begin{equation}
\label{eq:L0def}
{\cal L}^{(0)} \hat{\chi}_\mathrm{tot}  = -\frac{i}{\hbar}[\hat{H}_\mathrm{tot} \, , \hat{\chi}_\mathrm{tot}] + {\cal L}_{d,\mathrm{tot}}\hat{\chi}_\mathrm{tot} - {\cal L}^{(1)} \hat{\chi}_\mathrm{tot} .
\end{equation}
This means that the master equation reads 
\begin{equation}
\label{eq:QMEL0L1}
\dot{\hat{\chi}}_\mathrm{tot} = \left({\cal L}^{(0)} + {\cal L}^{(1)}  \right) \hat{\chi}_\mathrm{tot} \equiv {\cal L}  \hat{\chi}_\mathrm{tot} . 
\end{equation}

We now define the projection operator ${\cal P}$ by
\begin{equation}
\label{eq:Projdef}
{\cal P} \hat{X} = |\mathrm{vac}\rangle \langle \mathrm{vac}| \otimes \mathrm{Tr}_{-,3} \, \hat{X} ,
\end{equation}
where $|\mathrm{vac}\rangle$ denotes the vacuum states in modes $\hat{a}_-$ and $\hat{a}_3$, and $\mathrm{Tr}_{-,3}$ denotes tracing over the same modes. Applying the projection operator to the density matrix gives
\begin{equation}
\label{eq:Projdens}
{\cal P} \hat{\chi}_\mathrm{tot} = |\mathrm{vac}\rangle \langle \mathrm{vac}| \otimes \hat{\chi}_{+,m}  \  , \quad \hat{\chi}_{+,m} \equiv \mathrm{Tr}_{-,3} \, \hat{\chi}_\mathrm{tot} .
\end{equation}
The complement to the projection operator is defined as ${\cal Q} = 1 - {\cal P}$. With these definitions, we have the relations
\begin{eqnarray}
\label{eq:RelProj}
{\cal P} {\cal L}^{(1)} {\cal P} & = & 0 \\
{\cal Q} {\cal L}^{(1)} {\cal P} & = & {\cal L}^{(1)} {\cal P} \notag \\
{\cal P} {\cal L}^{(0)} {\cal P} & = & {\cal L}^{(0)} {\cal P} \notag \\
{\cal Q} {\cal L}^{(0)} {\cal P} & = & 0 \notag \\
{\cal Q} {\cal L}^{(0)} {\cal L}^{(1)} {\cal P} & = & {\cal L}^{(0)} {\cal L}^{(1)} {\cal P} , \notag
\end{eqnarray}
which will be needed below.

The master equation \eqref{eq:QMEL0L1} can now be expressed in terms of coupled equations for the projection ${\cal P} \hat{\chi}_\mathrm{tot}$ and its complement ${\cal Q} \hat{\chi}_\mathrm{tot}$:
\begin{eqnarray}
\label{eq:MEPrjCompl}
\frac{d}{dt} {\cal P}\hat{\chi}_\mathrm{tot} & = & {\cal P}{\cal L} {\cal P}\hat{\chi}_\mathrm{tot} + {\cal P}{\cal L} {\cal Q}\hat{\chi}_\mathrm{tot} \\
\frac{d}{dt} {\cal Q}\hat{\chi}_\mathrm{tot} & = & {\cal Q}{\cal L} {\cal P}\hat{\chi}_\mathrm{tot} + {\cal Q}{\cal L} {\cal Q}\hat{\chi}_\mathrm{tot} .
\end{eqnarray}
By formally solving the latter equation in the limit $t \rightarrow \infty$ where the solution does not depend on initial conditions, we get
\begin{equation}
\label{eq:Qsol}
{\cal Q}\hat{\chi}_\mathrm{tot}(t) = \int_0^\infty d \tau \, \ex^{{\cal Q} {\cal L} \tau} {\cal Q} {\cal L} {\cal P} \hat{\chi}_\mathrm{tot}(t - \tau).
\end{equation}
We then insert this into the equation for ${\cal P} \hat{\chi}_\mathrm{tot}$. By expanding to second order in ${\cal L}^{(1)}$ and using the relations \eqref{eq:RelProj}, we get
\begin{eqnarray}
\label{eq:sdfasdfasd}
\frac{d}{dt} {\cal P}\hat{\chi}_\mathrm{tot} & = & {\cal L}^{(0)} {\cal P}\hat{\chi}_\mathrm{tot} \\
& + & \int_0^\infty d \tau  {\cal P} {\cal L}^{(1)} \ex^{{\cal L}^{(0)} \tau} {\cal L}^{(1)} {\cal P}\hat{\chi}_\mathrm{tot}(t - \tau). \notag
\end{eqnarray}
This gives the following equations for the reduced density matrix $\hat{\chi}_{+,m}$:
\begin{eqnarray}
\label{eq:SME1}
& & \dot{\hat{\chi}}_{+,m} =  {\cal L}^{(0)}_{+,m} \hat{\chi}_{+,m} -  g_{+-}^2 \int_0^\infty d \tau \\
& & \times \, \Big\{ {\cal G}_{-}^{(0)}(\tau) \left[(\hat{c} + \hat{c}^\dagger ) \hat{a}_+^\dagger \, , \, \ex^{{\cal L}^{(0)}_{+,m} \tau} (\hat{c} + \hat{c}^\dagger ) \hat{a}_+ \hat{\chi}_{+,m}(t-\tau) \right] \notag \\
& & -  \, {\cal G}_{-}^{(0) \, \ast}(\tau) \left[(\hat{c} + \hat{c}^\dagger ) \hat{a}_+ \, , \, \ex^{{\cal L}^{(0)}_{+,m} \tau} \hat{\chi}_{+,m}(t-\tau) (\hat{c} + \hat{c}^\dagger ) \hat{a}_+^\dagger  \right] \notag   \\
& & - \, G_\uparrow^2 \int_0^\infty d \tau \notag \\
& & \times \, \Big\{ {\cal G}_{3}^{(0)}(\tau) \left[(\hat{c} + \hat{c}^\dagger ) \, , \, \ex^{{\cal L}^{(0)}_{+,m} \tau} (\hat{c} + \hat{c}^\dagger ) \hat{\chi}_{+,m}(t-\tau) \right] \notag \\
& & -  \, {\cal G}_{3}^{(0) \, \ast}(\tau) \left[(\hat{c} + \hat{c}^\dagger )  \, , \, \ex^{{\cal L}^{(0)}_{+,m} \tau} \hat{\chi}_{+,m}(t-\tau) (\hat{c} + \hat{c}^\dagger )  \right] \notag \Big\} .
\end{eqnarray}
We have defined the Green's functions
\begin{eqnarray}
\label{eq:Green}
{\cal G}_{-}^{(0)}(\tau) & = & \langle \hat{a}_-(\tau) \hat{a}_-^\dagger(0) \rangle^{(0)} \\ 
{\cal G}_{3}^{(0)}(\tau) & = & \langle \hat{a}_3(\tau) \hat{a}_3^\dagger(0) \rangle^{(0)}
\end{eqnarray}
where the superscript $(0)$ indicates that they are calculated with respect to the unperturbed Liouvillian ${\cal L}^{(0)}$. We have also defined
\begin{eqnarray}
\label{eq:L0plusmDef}
{\cal L}^{(0)}_{+,m} \hat{X} & = & -i \Big[- \Delta_+ \hat{a}^\dagger_+ \hat{a}_+ + \omega_m \hat{c}^\dagger \hat{c}  \\ 
 & + & G_\downarrow (\hat{c} + \hat{c}^\dagger ) (\hat{a}_+ + \hat{a}_+^\dagger) + g_{++} (\hat{c} + \hat{c}^\dagger ) \hat{a}^\dagger_+ \hat{a}_+ \, , \, \hat{X} \Big] \notag \\
& + & \left( \kappa_+ {\cal D}[\hat{a}_+]  + \gamma_{\downarrow, \therm}  {\cal D}[\hat{c}] + \gamma_{\uparrow, \therm} {\cal D}[\hat{c}^\dagger] \right) \hat{X} . \notag 
\end{eqnarray}

The Green's functions can be calculated either by using the quantum regression theorem \cite{Carmichael1993Book} or from quantum Langevin equations \cite{Clerk2010RMP}. The result is
\begin{eqnarray}
\label{eq:Greensolutions}
{\cal G}_{-}^{(0)}(\tau) & = & \ex^{-\left(\kappa_-/2 + i\omega_m\right)\tau} \\ 
{\cal G}_{3}^{(0)}(\tau) & = & \ex^{-\left(\kappa_3/2 - i\omega_m \right)\tau} .
\end{eqnarray}
These functions will suppress the $\tau$-integrands in \eqref{eq:SME1} for times $\tau > 1/\kappa_-, 1/\kappa_3$. For $\tau$ smaller than this, we can make the approximation that the evolution operator $\ex^{{\cal L}^{(0)}_{+,m} \tau}$ only leads to free evolution of the operators $\hat{a}_+$ and $\hat{c}$. This is accurate as long as we assume $\kappa_+, \gamma_{\downarrow, \therm}, \gamma_{\uparrow, \therm}, G_\downarrow^2/\omega_m , g^2_{++}/\omega_m \ll \kappa_-, \kappa_3$. Also, we exploit the fact that to zeroth order in ${\cal L}^{(1)}$, we have $\ex^{{\cal L}^{(0)}_{+,m} \tau}\hat{\chi}_{+,m}(t - \tau) = \hat{\chi}_{+,m}(t)$. This gives
\begin{eqnarray}
\label{eq:evolve}
& & \ex^{{\cal L}^{(0)}_{+,m} \tau} (\hat{c} + \hat{c}^\dagger ) \hat{a}_+ \hat{\chi}_{+,m}(t-\tau) \\ & & \quad \approx \left(\ex^{i \omega_m \tau} \hat{c} + \ex^{-i \omega_m \tau} \hat{c}^\dagger \right) \ex^{-i \Delta_+ \tau} \hat{a}_+ \hat{\chi}_{+,m}(t) , \notag
\end{eqnarray}
and similarly for the other terms of this type.

The terms proportional to $G_\downarrow$ and $g_{++}$ in \eqref{eq:L0plusmDef} are off-resonant, since $\Delta_+ \approx - 2 \omega_m$. Many of the terms in \eqref{eq:SME1} originating from the projection procedure are also off-resonant, but with smaller prefactors ($G_\uparrow^2/\kappa_3$, $G_\uparrow^2/\omega_m$, $g_{+-}^2/\kappa_-$, $g_{+-}^2/\omega_m$). These small off-resonant terms are suppressed due to the large mechanical frequency $\omega_m$ and we neglect them in the following. The master equation then turns into
\begin{eqnarray}
\label{eq:SME2}
\dot{\hat{\chi}}_{+,m} & = & \tilde{{\cal L}}^{(0)}_{+,m} \hat{\chi}_{+,m} - i \Lambda \left[ \hat{c}^\dagger \hat{c} \, \hat{a}_+^\dagger \hat{a}_+ \, , \, \hat{\chi}_{+,m} \right] \\
& + & \left(\Gamma {\cal D}[\hat{c}^\dagger \hat{a}_+] + \Gamma_\downarrow {\cal D}[\hat{c} \, \hat{a}_+] \right) \hat{\chi}_{+,m} , \notag
\end{eqnarray}
where $\Gamma = 4 g_{+-}^2/\kappa_-$, $\Gamma_\downarrow = (g_{+-}/(2 \omega_m))^2 \kappa_-$, $\Lambda = g_{+-}^2/(2\omega_m)$, and
\begin{eqnarray}
\label{eq:LtildeDef}
\tilde{{\cal L}}^{(0)}_{+,m} \hat{X} & = & -i \Big[- \Delta_+ \hat{a}^\dagger_+ \hat{a}_+ + \bar{\omega}_m \hat{c}^\dagger \hat{c}  \\ 
 & + & G_\downarrow (\hat{c} + \hat{c}^\dagger ) (\hat{a}_+ + \hat{a}_+^\dagger) + g_{++} (\hat{c} + \hat{c}^\dagger ) \hat{a}^\dagger_+ \hat{a}_+ \, , \, \hat{X} \Big] \notag \\
& + & \left( \kappa_+ {\cal D}[\hat{a}_+]  + \bar{\gamma}_{\downarrow}  {\cal D}[\hat{c}] + \bar{\gamma}_{\uparrow} {\cal D}[\hat{c}^\dagger] \right) \hat{X} . \notag
\end{eqnarray}
The last equation differs from \eqref{eq:L0plusmDef} in that the mechanical frequency and dissipation rates have been renormalized:
\begin{eqnarray}
\label{eq:Renorm}
\bar{\omega}_m & = & \omega_m + \frac{G_\uparrow^2}{2\omega_m} \\
\bar{\gamma}_\downarrow & = & \gamma_{\downarrow, \therm} + \left(\frac{G_\uparrow}{2 \omega_m}\right)^2 \kappa_3 \notag \\
\bar{\gamma}_\uparrow & = & \gamma_{\uparrow, \therm} + \frac{4 G_\uparrow^2}{\kappa_3} .  \notag
\end{eqnarray}
The dissipator ${\cal D}[\hat{c} \, \hat{a}_+]$ in \eqref{eq:SME2} annihilates both a photon in the plus mode and a phonon. However, since $\Gamma_\downarrow \ll \Gamma$, this process is suppressed and we can neglect this term. The cross-Kerr term proportional to $\Lambda$ will also not be of importance, since $\Lambda \ll \Gamma$. These considerations lead to the master equation
\begin{eqnarray}
\label{eq:SME3}
\dot{\hat{\chi}}_{+,m} & = & \tilde{{\cal L}}^{(0)}_{+,m} \hat{\chi}_{+,m} + \Gamma {\cal D}[\hat{c}^\dagger \hat{a}_+]  \hat{\chi}_{+,m} .
\end{eqnarray}

To conclude, we see that the remnants of the mode $\hat{a}_-$ is to provide a decay channel for photons in $\hat{a}_+$, but one where the destruction of a photon is associated with the creation of a phonon. The effect of coupling to the mode $\hat{a}_3$ is simply to renormalize the parameters in \eqref{eq:Renorm} in such a way that $\bar{\gamma}_\uparrow \gg \bar{\gamma}_\downarrow$.

\subsection{4.2. The unitary transformation $\hat{U}$}

The Liouvillian $\tilde{{\cal L}}^{(0)}_{+,m}$ is identical to that of a standard optomechanical system where the optical mode is coherently driven \cite{Borkje2013PRL_SM}. The bilinear interaction term proportional to $G_\downarrow$ gives normal modes that are linear combination of photons and phonons. In our off-resonant case $\Delta_+ \sim -2 \omega_m$ and with $G_\downarrow/\omega_m \ll 1$, the mixing between photons and phonons is weak and the transformation to normal modes can be expanded to second order in $G_\downarrow/\omega_m$. We define $\hat{U} = \ex^{-\hat{\eta}}$ with
\begin{eqnarray}
\label{eq:etadef}
\hat{\eta} & = & \frac{G_\downarrow}{\Delta_+ + \bar{\omega}_m} \left(\hat{a}_+^\dagger \hat{c} - \hat{c}^\dagger \hat{a}_+\right) \\
& + & \frac{G_\downarrow}{\Delta_+ - \bar{\omega}_m} \left(\hat{a}_+^\dagger \hat{c}^\dagger - \hat{c} \hat{a}_+\right) \notag \\
& + & \frac{G_\downarrow^2 \bar{\omega}_m}{2 \Delta_+ \left(\Delta_+^2 - \bar{\omega}_m^2 \right)} \left(\hat{a}_+^{\dagger \, 2} - \hat{a}_+^2\right) \notag \\
& - & \frac{G_\downarrow^2 \Delta_+}{2 \bar{\omega}_m \left(\Delta_+^2 - \bar{\omega}_m^2 \right)} \left(\hat{c}^{\dagger \, 2} - \hat{c}^2\right) \notag
\end{eqnarray}
and the transformed density matrix 
\begin{equation}
\label{eq:tilderhodef}
\hat{\rho} = \hat{U}^\dagger \hat{\chi}_{+,m} \hat{U} .
\end{equation}
From \eqref{eq:SME3}, we can derive the master equation that determines $\tilde{\rho}$. To second order in $G_\downarrow/\omega_m$, we have
\begin{eqnarray}
\label{eq:transf}
\hat{U}^\dagger \hat{a}_+ \hat{U} & = & \hat{a}_+ + \left[\hat{\eta} \, , \, \hat{a}_+ \right] + \frac{1}{2} \left[\hat{\eta} \, , \left[\hat{\eta} \, , \, \hat{a}_+ \right] \right]  
\end{eqnarray}
and similarly for $\hat{c}$. Using this and neglecting several small terms and off-resonant terms that are suppressed, we arrive at the master equation
\begin{eqnarray}
\label{eq:SME4}
\dot{\hat{\rho}} & = & - i \left[- \tilde{\Delta}_+\hat{a}^\dagger_+ \hat{a}_+ + \tilde{\omega}_m \hat{c}^\dagger \hat{c} + \tilde{g} \left(\hat{a}_+^\dagger \hat{c}^2 + \hat{c}^{\dagger \, 2} \hat{a}_+ \right) \, , \, \hat{\rho} \right] \notag \\
& + & \left(\Gamma {\cal D}[\hat{c}^\dagger \hat{a}_+] + \kappa_+ {\cal D}[\hat{a}_+]  + \gamma_{\downarrow}  {\cal D}[\hat{c}] + \gamma_{\uparrow} {\cal D}[\hat{c}^\dagger] \right) \hat{\rho} .
\end{eqnarray}
Here, we have defined
\begin{eqnarray}
\label{eq:TildeDefs}
\tilde{\omega}_m & = & \bar{\omega}_m + \frac{2 G_\downarrow^2 \Delta_+}{\Delta_+^2 - \bar{\omega}_m^2} \\
\tilde{\Delta}_+ & = & \Delta_+ - \frac{2 G_\downarrow^2 \bar{\omega}_m}{\Delta_+^2 - \bar{\omega}_m^2} \\
\tilde{g} & = & \frac{g_{++} G_\downarrow}{\Delta_+ + \bar{\omega}_m}
\end{eqnarray}
The exact value of $\Delta_+$ is found by requiring $\tilde{\Delta}_+ = -2 \tilde{\omega}_m$, which gives
\begin{equation}
\label{eq:DeltaPlus}
\Delta_+ = -2\omega_m \left[1 - \frac{5}{3}\left(\frac{G_\downarrow}{\omega_m}\right)^2 + \frac{1}{2}\left(\frac{G_\uparrow}{\omega_m}\right)^2 \right].
\end{equation}
In practice, the accuracy of the detuning need only be smaller than $\Gamma$. 

The final step is to move to rotating frames for both modes, which is done by another transformation
\begin{equation}
\label{eq:rhoDef}
\tilde{\rho} = \hat{V}^\dagger \hat{\rho} \hat{V}
\end{equation}
with 
\begin{equation}
\label{eq:Vdef}
\hat{V} = \ex^{-i \tilde{\omega}_m \left(2 \hat{a}_+^\dagger \hat{a}_+ + \hat{c}^\dagger \hat{c} \right) t} .
\end{equation}
When renaming $\tilde{\rho} \rightarrow \hat{\rho}$, we then arrive at the generic model defined in Eqs.~(1) and (2) of the main article.

\section{5. Approximate analytical solution to optomechanical model}

In Sec.~1, we presented analytical expressions for the steady state density matrix of the generic model. We now apply these to the  optomechanical example to estimate the occupation probability in the $n=1$ phonon state. This requires that we transform back to the basis in terms of photons and phonons, i.e.~we want the density matrix
\begin{equation}
\label{eq:transformBack}
\hat{\chi}_{+,m} = \hat{U} \hat{V} \hat{\rho} \hat{V}^\dagger \hat{U}^\dagger . 
\end{equation}
First, we note that for the density matrix in Eq.~\eqref{eq:ansatzrho}, we have $\hat{V} \hat{\rho} \hat{V}^\dagger = \hat{\rho}$. We then expand to second order in $G_\downarrow/\omega_m$ as before, giving
\begin{equation}
\label{eq:chiApprox}
\hat{\chi}_{+,m} = \hat{\rho} - \left[\hat{\eta} \, , \, \hat{\rho} \right] + \frac{1}{2} \left[\hat{\eta} \, , \left[\hat{\eta} \, , \, \hat{\rho} \right] \right] .
\end{equation}
This gives small corrections compared to $\hat{\rho}$ both in the diagonal and the off-diagonal elements in the Fock basis. For calculating occupation probabilities, we only need the diagonal elements. 

\begin{figure}[h]
 \centering
 \includegraphics[width=0.99\columnwidth,trim=0cm 0cm 0cm 0cm]{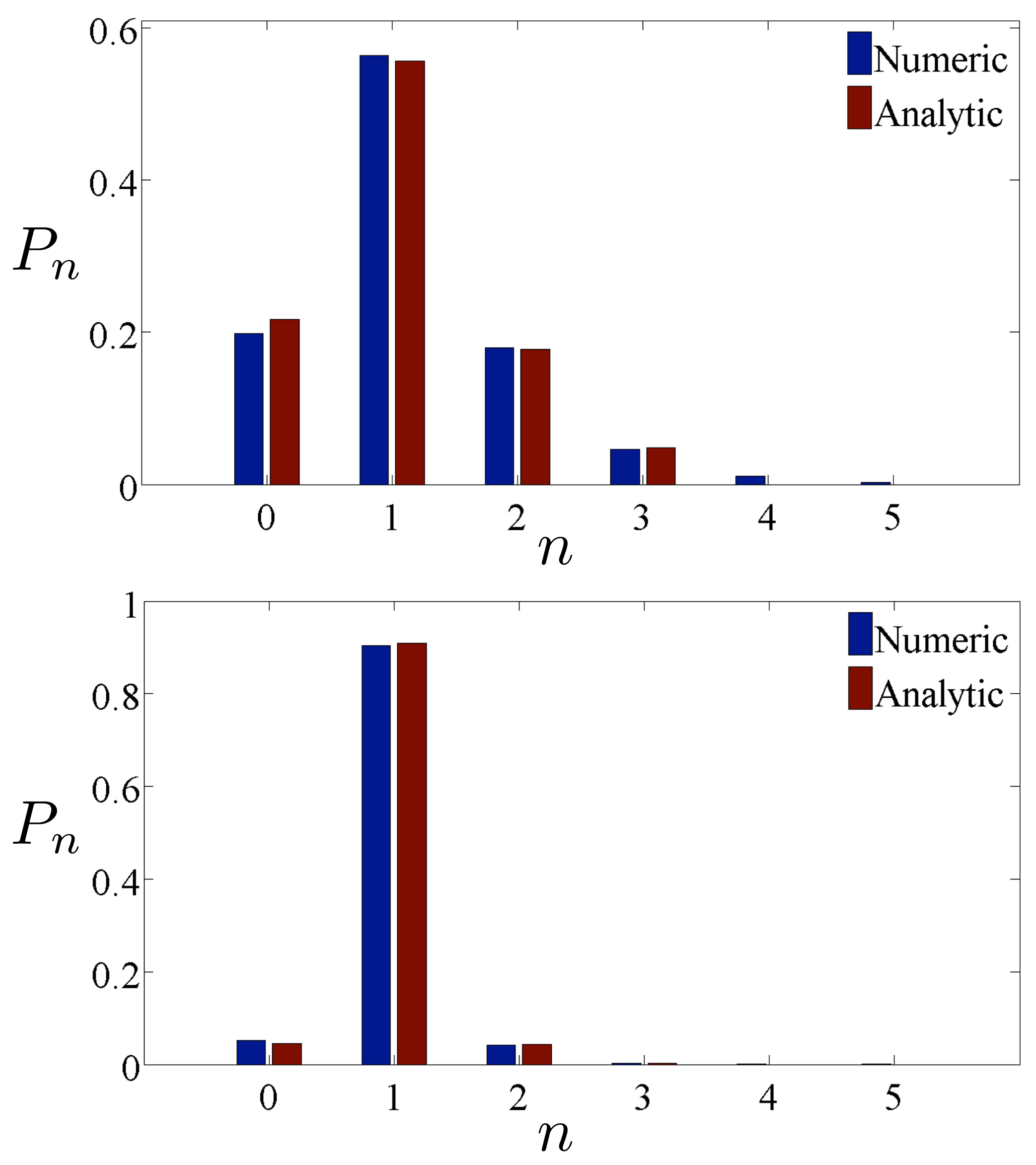}
\caption{(color online). Comparison of the numerical and analytical results for the phonon occupation probabilities $P_n$. The parameters used in the upper panel are the same as in Fig.~3(a,b) of the main article. In the lower panel, we used the same parameters as in Fig.~3(c,d).}
\label{fig:OccProb}
\end{figure}

We define the coefficients $\chi_{kk',nn'}$ by
\begin{equation}
\label{eq:coeffdef}
\hat{\chi}_{+,m} = \sum_{kk'} \sum_{nn'} \chi_{kk',nn'} |k \rangle \langle k' | \otimes |n \rangle \langle n'| , 
\end{equation}
where $k,k'$ refers to photon Fock states in the mode $\hat{a}_+$ and $n,n'$ to phonon Fock states in mode $\hat{c}$. We assume that we are in the regime where $\rho_{00,1}$, as defined in Eq.~\eqref{eq:ansatzrho}, is large compared to the other matrix elements. This allows us to simplify \eqref{eq:chiApprox} by only including the corrections that are proportional to $\rho_{00,1}$. The diagonal elements that get nonnegligible corrections from the reverse transform $\hat{U}^{-1} = \hat{U}^\dagger$ are then
\begin{eqnarray}
\label{eq:chiApproxDiag}
\hat{\chi}_{00,11} & = & \rho_{00,1} \left[1 - \left(\frac{G_\downarrow}{\Delta_+ + \bar{\omega}_m}\right)^2 -   \left(\frac{G_\downarrow}{\Delta_+ - \bar{\omega}_m}\right)^2 \right] \quad \\
\hat{\chi}_{11,00} & = & \rho_{11,0} + \rho_{00,1} \left(\frac{G_\downarrow}{\Delta_+ + \bar{\omega}_m}\right)^2 \\
\hat{\chi}_{11,22} & = & \rho_{00,1} \left(\frac{G_\downarrow}{\Delta_+ - \bar{\omega}_m}\right)^2 , 
\end{eqnarray}
whereas the other diagonal elements are unchanged, i.e.~$\chi_{kk,nn} = \rho_{kk,n}$. The phonon Fock state occupation probabilities are given by 
\begin{equation}
\label{eq:PhononProbs}
P_n = \sum_k \chi_{kk,nn} .
\end{equation}
In Fig.~\ref{fig:OccProb}, we compare these analytical results to the results from the numerical simulations. We use the same parameters as in Fig.~3 of the main article and find good agreement. 

\section{6. Definition of Wigner distribution and numerical results}

The Wigner quasi-probability distribution is defined as \cite{Davidovitch2000}
\begin{equation}
\label{eq:WignerDef}
W(q,p) = \frac{1}{2\pi} \int d q' \, \ex^{ipq'} \left\langle q - \frac{q'}{2} \right| \hat{\chi}_m \left| q + \frac{q'}{2} \right\rangle ,
\end{equation}
where $|q \rangle$ is an eigenstate of the dimensionless position operator
\begin{equation}
\label{eq:DimPos}
\hat{q} = \frac{1}{\sqrt{2}} \left(\hat{c} + \hat{c}^\dagger \right),
\end{equation}
meaning that $\hat{q}|q\rangle = q|q\rangle$. The definition of the Wigner distribution used here is normalized such that
\begin{equation}
\label{eq:WignerNorm}
\int_{-\infty}^\infty dq \int_{-\infty}^\infty dp \, W(q,p) = 1
\end{equation}
and bounded according to $|W(q,p)| \leq 1/\pi$.
  
For a Fock state $\hat{\chi}_m = |n \rangle \langle n |$, the Wigner distribution becomes
\begin{equation}
\label{eq:Wn}
W_n(q,p) = \frac{(-1)^n}{\pi} L_n\left[2(q^2 + p^2)\right] \ex^{-(q^2 + p^2)},
\end{equation}
where $L_n[\cdot]$ is the Laguerre polynomial of degree $n$. Specifically, for the single phonon Fock state $\hat{\chi}_m = |1 \rangle \langle 1 |$, we get
\begin{equation}
\label{eq:W1}
W_1(q,p) = \frac{1}{\pi} \left[2(q^2 + p^2) - 1\right] \ex^{-(q^2 + p^2)}.
\end{equation}
We see that this is negative in the region of phase space where $q^2 + p^2 \leq 1/2$, and that it is maximally negative ($-1/\pi$) at the origin $q=p=0$.  

\begin{figure}[h]
 \centering
 \includegraphics[width=0.99\columnwidth,trim=0cm 0cm 0cm 0cm]{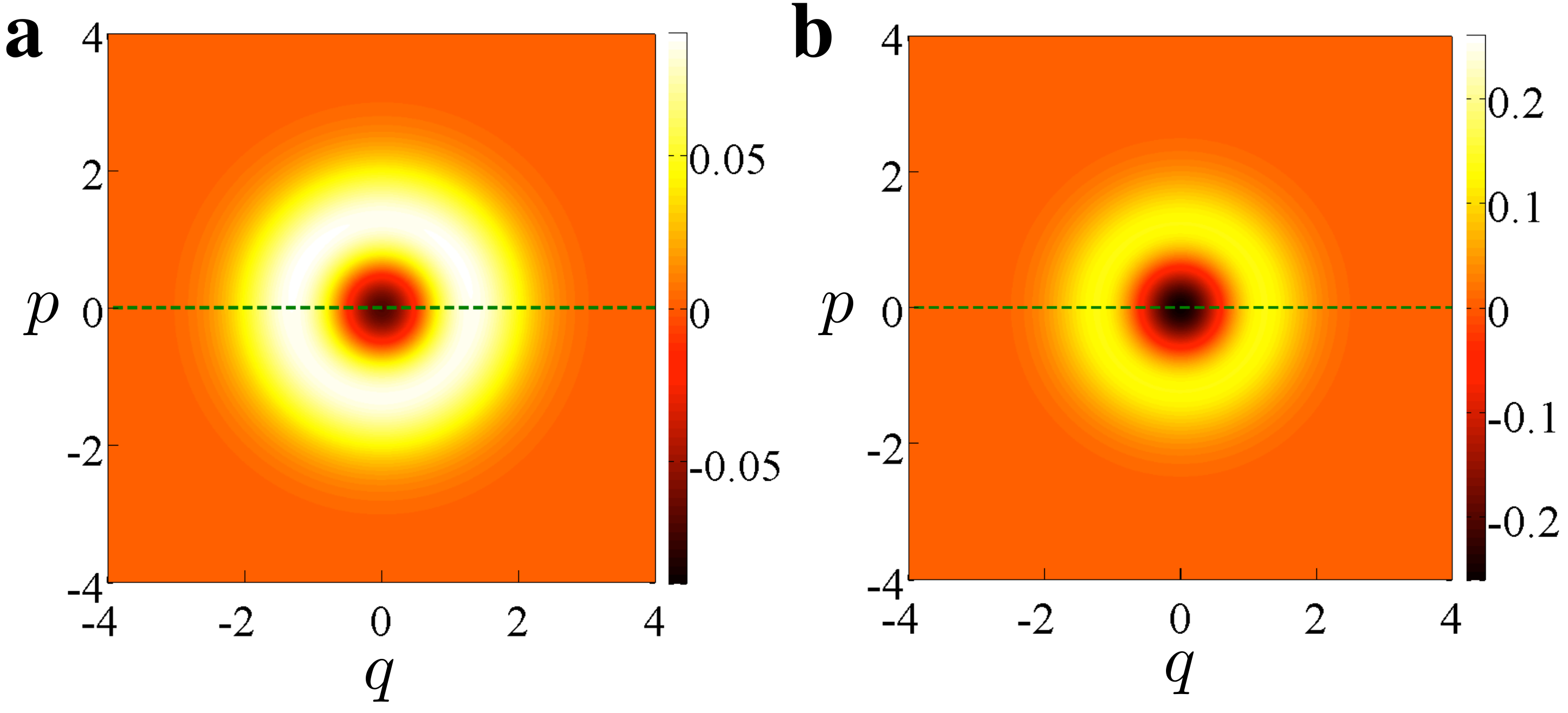}
\caption{(color online). Density plot of the Wigner distribution $W(q,p)$ for the parameters used in the main article. In (a), we used the parameters in Fig.~3(a,b). In (b), we used the same parameters as in Fig.~3(c,d). The dashed line at $p = 0$ corresponds to the plots in Figs.3(b,d).}
\label{fig:Wig3D}
\end{figure}

In Fig.~\ref{fig:Wig3D}, we plot the Wigner distribution as derived from our numerical simulations on the optomechanical model. We use the same parameters as in Fig.~3 of the main article. We observe that the Wigner distributions are rotationally symmetric around the center $q = p = 0$, and that there is a region of negativity at the center for both parameter sets.

\end{document}